\documentclass[useAMS,usenatbib]{mnras}

\usepackage[dvipsnames]{xcolor}
\usepackage{ragged2e}
\usepackage{tikz,hyperref}
\usepackage[normalem]{ulem}
\usepackage{enumitem}
\definecolor{lime}{HTML}{A6CE39}
\DeclareRobustCommand{\orcidicon}{
	\begin{tikzpicture}
	\draw[lime, fill=lime] (0,0) 
	circle [radius=0.13] 
	node[white] {{\fontfamily{qag}\selectfont \tiny ID}};
	\draw[white, fill=white] (-0.0625,0.095) 
	circle [radius=0.007];
	\end{tikzpicture}
	\hspace{-2mm}
}

\foreach \x in {A, ..., Z}{\expandafter\xdef\csname orcid\x\endcsname{\noexpand\href{https://orcid.org/\csname orcidauthor\x\endcsname}
			{\noexpand\orcidicon}}
}

\usepackage{graphicx}
\usepackage{supertabular}
\usepackage{amssymb,amsmath}
\usepackage[export]{adjustbox}
\usepackage{rotating}
\usepackage{soul}
\newcommand{\heiiwr}{He\,{\sc ii}\,$\lambda4686$}

\newcommand{\ewhb}{EW(H$\beta$)}

\newcommand{\msol}{M$_{\odot}$}
\newcommand{\msolyr}{M$_{\odot}$\,yr$^{-1}$}

\newcommand{\hi}{H\,{\small{\sc i}}}
\newcommand{\hii}{H\,{\small{\sc ii}}}
\newcommand{\ha}{H$\alpha$}
\newcommand{\hb}{H$\beta$}

\newcommand{\Av}{$A_{\rm V}$}
\newcommand{\Rv}{$R_{\rm V}$}

\newcommand{\cliii}{[Cl\,{\sc iii}]}
\newcommand{\ariv}{[Ar\,{\sc iv}]}

\newcommand{\feiii}{[Fe\,{\sc iii}]\,$\lambda4658$}
\newcommand{\Mgii}{Mg\,{\sc ii}}
\newcommand{\Nai}{Na\,{\sc i}}

\newcommand{\oiii}{[O{\,\scshape iii]}}
\newcommand{\oiiia}{[O{\,\scshape iii]}\,$\lambda4363$}

\newcommand{\sii}{[S{\,\scshape ii]}}

\newcommand{\siii}{[S{\,\scshape iii]}}
\newcommand{\siiia}{[S{\,\scshape iii]}\,$\lambda6312$}
\newcommand{\siiib}{[S{\,\scshape iii]}\,$\lambda9069$}
\newcommand{\nii}{[N{\,\scshape ii}]}
\newcommand{\niia}{[N{\,\scshape ii}]\,$\lambda5755$}

\title[Nebular abundance in the Cartwheel galaxy]{
Nebular abundance gradient in the Cartwheel galaxy using MUSE data
}
\author[Javier Zaragoza-Cardiel et al.]{Javier Zaragoza-Cardiel\thanks{Email: javier.zaragoza@inaoep.mx}$^{1,2\orcidA{}}$, 
V. Mauricio A. G\'omez-Gonz\'alez$^{3\orcidB{}}$, Divakara Mayya$^{1\orcidC{}}$, \newauthor and  Gerardo Ramos-Larios$^{4,5\orcidD{}}$
\\
$^{1}$Instituto Nacional de Astrof{\'\i}sica, \'Optica y Electr\'onica, Luis Enrique Erro 1, Tonantzintla 72840, Puebla, Mexico\\
$^{2}$Consejo Nacional de Ciencia y Tecnolog\'ia, Av. Insurgentes Sur 1582, 03940,  Mexico City, Mexico\\
$^{3}$Institute for Physics and Astronomy, Universit\"{a}t Potsdam, Karl-Liebknecht-Str. 24/25, D-14476 Potsdam, Germany\\
$^{4}$Instituto de Astronom\'\i a y Meteorolog\'\i a, CUCEI, Univ.\ de Guadalajara, Av.\ Vallarta 2602, Arcos Vallarta, 44130 Guadalajara, Mexico\\
$^{5}$CUCEI, Universidad de Guadalajara, Blvd. Marcelino Garc\'\i a Barrag\'an 1421, 44430, Guadalajara, Jalisco, Mexico \\
}

\begin{document}

\maketitle

\begin{abstract}
We here present the results from a detailed analysis of nebular abundances of commonly
observed ions in the collisional ring galaxy Cartwheel using the Very Large Telescope
(VLT) Multi-Unit Spectroscopic Explorer (MUSE) dataset. 
The analysis includes 221 \hii\ regions in the star-forming ring, in addition
to 40 relatively fainter \ha-emitting regions in the spokes, disk and the inner ring. 
The ionic abundances of He, N, O and Fe are obtained using the direct method (DM) for 9, 20, 20, and 17 ring \hii\ regions, respectively,
where  the S$^{++}$ temperature-sensitive line is detected.
For the rest of the regions, including all the nebulae between the inner and the outer ring, we obtained O abundances using the strong-line method (SLM).
 The ring regions have a median $12+\log\rm{\frac{O}{H}}$=8.19$\pm$0.15,  $\log\rm{\frac{N}{O}}=-$1.57$\pm$0.09 and $\log\rm{\frac{Fe}{O}}=-$2.24$\pm$0.09 using the DM. 
Within the range of O abundances seen in the Cartwheel,
the N/O and Fe/O values decrease proportionately with increasing O, suggesting local enrichment of O without corresponding enrichment  of primary N and Fe.
The O abundances of the disk \hii\ regions obtained using the SLM show a 
well-defined radial gradient. The mean O abundance of the ring \hii\ regions is lower by $\sim$0.1~dex as compared to the extrapolation of the radial gradient. The observed trends suggest the preservation of the pre-collisional abundance gradient, displacement of most of the processed elements to the ring, as predicted by the recent simulation by \cite{Renaud2018}, and post-collisional infall of metal-poor gas in the ring.

\end{abstract}


\section{Introduction}

The Cartwheel (a.k.a ESO\,350-40 or A0035-324) is considered the archetype of collisional ring galaxies \citep{1996FCPh...16..111A, 2010MNRAS.403.1516S}.
These type of galaxies are formed as a result of a compact galaxy plunging through a massive gas-rich disk galaxy close to its center and almost perpendicular to it \citep{1976ApJ...209..382L} and are characterized by a ring that harbours a chain of star-forming knots \citep{Marston1995, Higdon1995}.
In classical models of ring galaxies, the star formation (SF) is triggered by a radially expanding density wave that leaves behind the older generations of stars towards the center \citep{Marcum1992, Appleton1997}. 
In a recent hydrodynamical simulation that treats SF at parsec scales, \citet{Renaud2018} find  
that the wave drags the gas and recently formed stars along with it to the present location of the ring.

The chemical properties of the disk and the star-forming ring in the two scenarios presented above are distinctly different. The products of nucleosynthesis expelled from stars are accumulated in the ring in the scenario presented by \citet{Renaud2018}, whereas they stay at the location where the stars formed in the classical scenario \citep[see e.g.][]{Korchagin1999}. Data on chemical abundances on ring galaxies are scarce and do not cover the spatial extent required to test the two scenarios. The most extensive study was carried out by \citet{Bransford1998}, who obtained O and N abundances from long-slit spectroscopic data for 28 \hii\ regions in the star-forming ring of eight northern ring galaxies. Their empirically determined O abundances were uniform over the sample, thus ruling out prompt local enrichment. More recently, \cite{2021NGC922} found a radial abundance gradient in the C-shaped ring galaxy NGC\,922, using empirically-derived O abundances of eight \hii\ regions. For the Cartwheel, \citet{Fosbury1977} measured an O abundance of 12+log(O/H)$\sim$8.0 and log(N/O)=$-1.5$, using optical spectra of three of the brightest ring \hii\ regions.

\begin{figure*}
\begin{centering}
\includegraphics[width=1\linewidth]{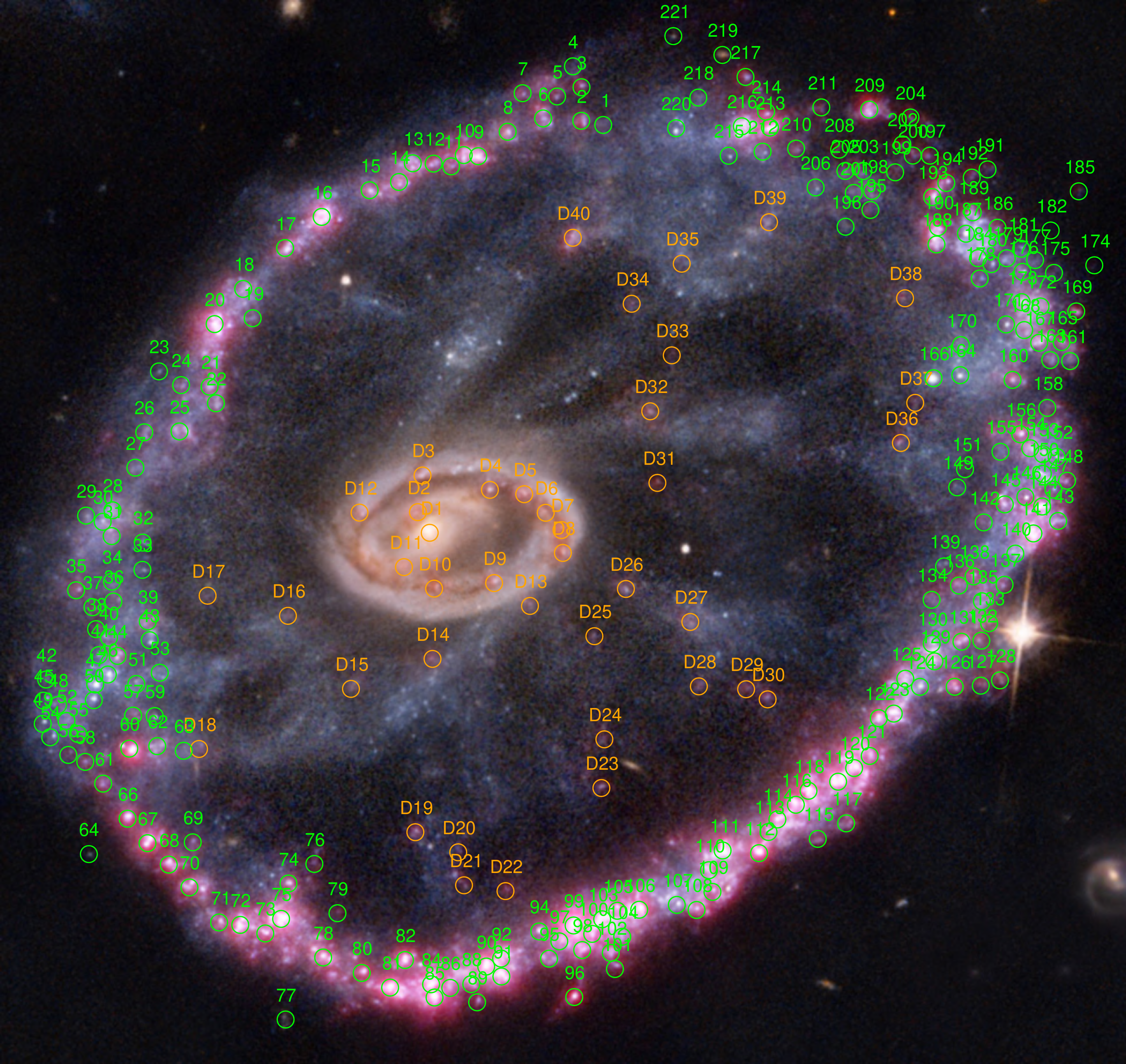}
\par\end{centering}
\caption{
 Colour-composite image of the Cartwheel galaxy formed using the HST/WFPC2 images (PSF=0.2~arcsec) 
in F814W, pseudo-green and F450W as red, green and blue components, respectively; and the \ha\ image constructed from MUSE data (PSF=0.6~arcsec) as a fourth reddish component. The \hii\ regions in the outer ring and interior to the outer ring are identified by green and orange circles, respectively. The circles are of 0.6~arcsec radius, which corresponds to 370~pc at the distance of the Cartwheel.
}
\label{fig:muse_image}
\end{figure*}

The Cartwheel has been the target of Multi-Unit Spectroscopic Explorer
(MUSE) observations at the Very Large Telescope (VLT),
that covers the entire optical extent of the galaxy at the seeing-limited spatial resolution of $\sim$0.6~arcsec.
The spectra span a rest wavelength range from $\sim$4610 to $\sim$9075~\AA,
which includes emission lines from different ionic species suitable for the determination of  abundances of commonly reported elements covering the entire galaxy. At the distance of the Cartwheel (128~Mpc using the Hubble constant of 71~km\,s$^{-1}$\,Mpc$^{-1}$ and velocity=9050 km\,s$^{-1}$), MUSE spectra are available at physical scales of $\sim$370~pc. 
On the \ha\ image constructed using this dataset, we have identified 221 individual \hii\ regions in the star-forming outer ring and 40 regions in the apparently quiescent inner zones, including around 10 regions in the inner ring. These regions are identified in a colour-composite image formed using this \ha\ image in Fig.~\ref{fig:muse_image}. This dataset offers an excellent opportunity to confront the predictions of the two scenarios of ring formation discussed above.

This article is organised as follows. In Section~2 we describe the dataset, extraction of individual spectra and details of measurement of line fluxes. Techniques followed for the determination of abundances are described in Section~3. Detailed discussions based on the reported abundances are presented in the context of the scenarios for the formation of ring galaxies in Section~4. We present our conclusions in Section~5.

\begin{figure*}
\begin{center}
\includegraphics[width=1\linewidth]{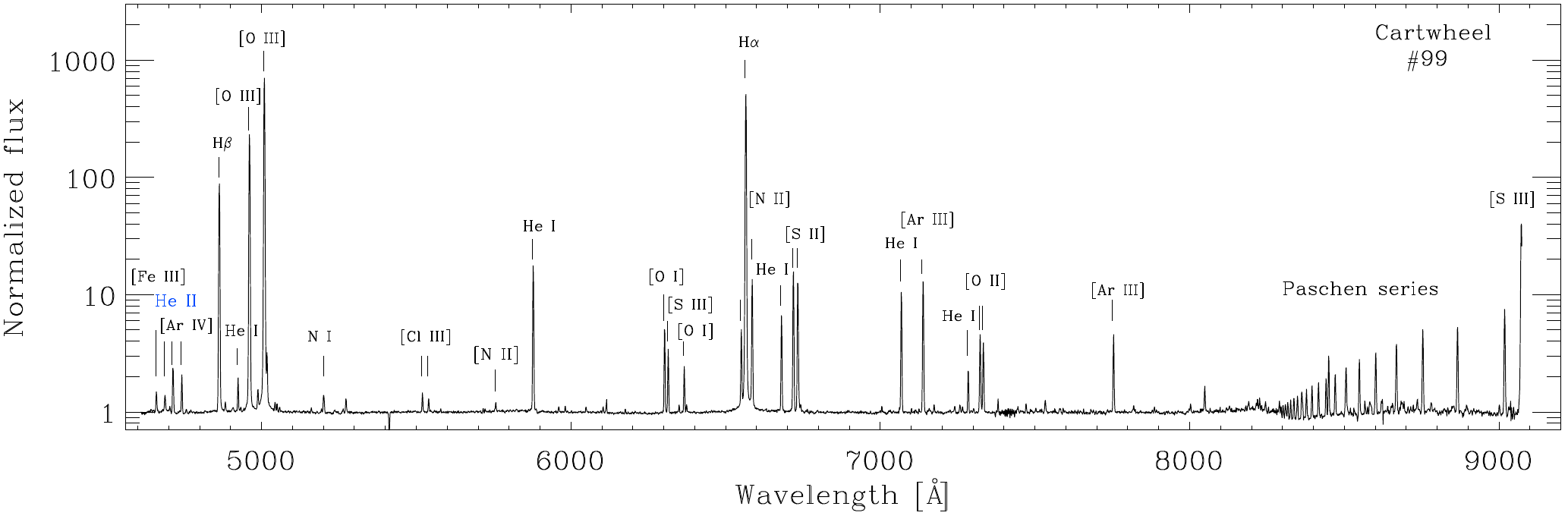}
\caption{VLT MUSE redshift corrected spectrum of cluster \#99 in the Cartwheel.
The most common nebular lines are indicated.
The spectrum is shown normalized to the best fit continuum spectrum.}
\label{fig:MUSE_spec}
\end{center}
\end{figure*}

\section{Spectroscopic data}
\subsection{VLT/MUSE observations}

MUSE is a panoramic integral-field spectrograph at the 8~m VLT of the European Southern Observatory (ESO)\footnote{\url{http://muse-vlt.eu/science/}}, operating in the optical wavelength range from 4750 to 9351~\AA\, with a spatial sampling of 0.2 arcsec\,pixel$^{-1}$, a spectral sampling of 1.25~\AA\,pixel$^{-1}$ and a spectral resolution $\sim$3~\AA\ \citep{Bacon2010}.
The public MUSE data cube covering the entire Cartwheel galaxy was retrieved from the ESO Science Archive Facility\footnote{\url{http://archive.eso.org/cms.html}}. The data cube was already flux and wavelength calibrated, and hence, ready for scientific exploitation. The observations were carried out during three runs on 2014 August 24 and 25 (PI: science verification, MUSE) with an effective exposure time of $\approx$4030~s.
The observations covered a field of view (FoV) of 2$\times$2~arcmin$^2$, which encloses the
entire star-forming ring of the Cartwheel, as can be seen in Fig.~\ref{fig:muse_image}.
Milky Way Galactic extinction in the Cartwheel galaxy line of sight is minimal, $A_{\rm{V}}=0.0285\thinspace \rm{mag}$  \citep{2011ApJ...737..103S}, so we do not correct for this effect.
As an example, we show in Fig.~\ref{fig:MUSE_spec} the MUSE rest-frame spectrum for the brightest \hii\ region in the Cartwheel, designated as region \#99 in this study. Prominent nebular lines are identified in the spectrum, that covers the \feiii\  and \siiib\ lines at the two extreme ends.

\subsection{MUSE data cube and astrometry}

We used the reduced 3-D data cube as a starting point in our analysis.
The tool QFitsView\footnote{\url{https://www.mpe.mpg.de/~ott/dpuser/qfitsview.html}}
\citep{Ott2012} was used for the extraction of images at
selected wavelengths, as well as spectra of selected regions.
A continuum image covering the entire spectral range of MUSE is prepared to facilitate the astrometric calibration of the image. We used the astrometrically calibrated
K-band image by \citet{Barway2020} to transform the MUSE and the Hubble Space Telescope (HST) images into the International Celestial Reference System (ICRS) as defined by
the coordinates of the point sources in the 2MASS catalog.
For this purpose, we identified point sources in the FoV of the MUSE
image that are present in the deep K-band image and/or the HST images.
We first astrometrised the HST images, which are then used to astrometrise
the MUSE images. The astrometric calibration was carried out
with the help of {\sc iraf}{\footnote{ IRAF is distributed by the National Optical Astronomy Observatories, which
are operated by the Association of Universities for Research in Astronomy,
Inc., under cooperative agreement with the National Science Foundation.}} tasks {\it ccmap} and {\it wregister}. Sixteen point sources were identified on the MUSE image which resulted
in the final absolute astrometric precision of 0.1~arcsec.

\subsection{Identifications of nebulae and spectral extraction}

As a first step, we generated a continuum-free \ha\ image from the MUSE datacube using
the scripts in QFitsView. For achieving this, we summed fluxes in a 20~\AA\ window centered on \ha\ at the redshifted wavelength of the nucleus of the Cartwheel (sum of pixel numbers between 1605 and 1620 in the wavelength axis) and subtracted a continuum image from it, which is obtained as average of two images 37.5~\AA\ away on either side of the \ha\ line.
On this continuum-free \ha\ image (the  reddish component in Fig.~\ref{fig:muse_image}), we visually identified 221 \hii\ regions in and around the outer ring\footnote{Among the 221 \ha-emitting regions, only 212 regions that have \hb\ line also in emission are retained in our final analysis.} and 40 regions in the region between the nucleus and the outer ring.
Spectra were extracted in apertures of 0.6~arcsec radius around all these visually identified regions. 

\citet{Higdon1995} had identified 29 star forming complexes in the outer ring based on an \ha\ image. The MUSE image we have used has a factor of 3 better seeing, allowing us to resolve each  complex into several \ha-emitting knots. At the resolution of the HST image ($\sim$0.2~arcsec=125~pc), most of our identified regions are associated with at least one star cluster that is responsible for its ionization. Thus, the selected ring regions are genuine \hii\ regions. On the other hand, the ionized zones selected in the inner disk cannot be associated to identifiable clusters. Furthermore, the extracted spectra of these disk regions showed clear signs for the presence of  stellar absorption at \hb, \Mgii\ and \Nai, that are in general broader than the nebular \hb\ line. 
They are likely \hii\ regions ionized by late B-type stars in relatively older fainter clusters, below the detection limit of the HST images. We will address elsewhere the nature of the ionization of these sources.

 Analysis of nebular abundances requires determination of flux ratios of nebular lines relative to the flux of the \hb\ line, which in the inner disk regions is clearly affected by the presence of the underlying absorption feature. 
We fitted the continuum with population synthesis models using the GIST pipeline  \citep{2019A&A...628A.117B}, which makes use of pPXF \citep{2017MNRAS.466..798C} and the MILES library \citep{2010MNRAS.404.1639V} to fit the continuum and the absorption lines.
 The population synthesis was performed over the 4750-7000~\AA\ spectral range, where the MUSE spectral resolution ranges from 3~\AA\ at 4750~\AA\ to 2.5~\AA\ at 7000~\AA\ \citep{2017A&A...608A...1B}. Therefore, we are  able to use the MILES library which has a spectral resolution of 2.5~\AA\ and covers the 3540-7400~\AA\ spectral range \citep{2010MNRAS.404.1639V}. None of the nebular lines outside of the fitted spectral range are affected by the underlying absorption feature, and hence the flux ratios measured on the observed spectrum are equally good.

\begin{figure*}
\begin{centering}
\includegraphics[width=0.5\linewidth]{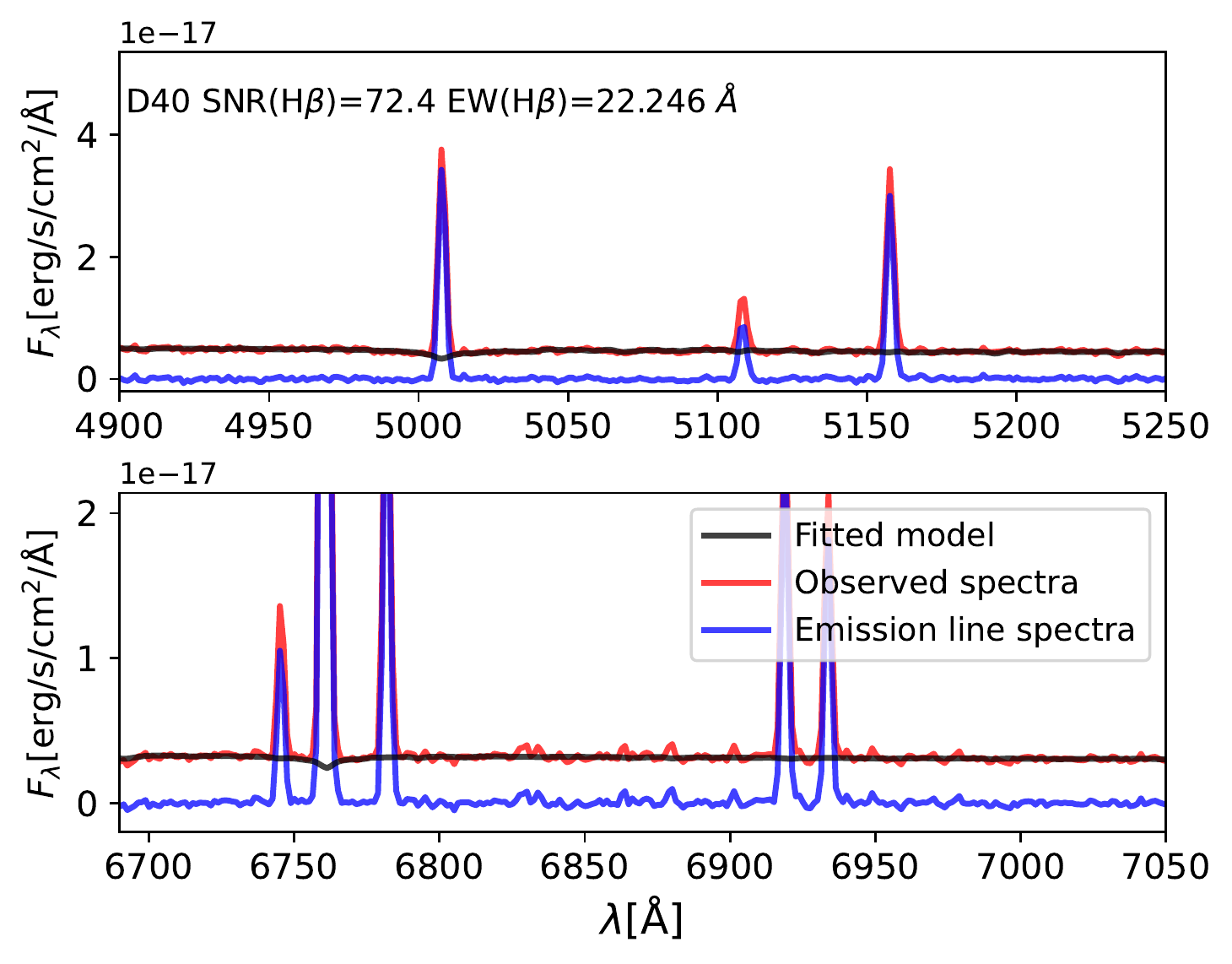}~
\includegraphics[width=0.5\linewidth]{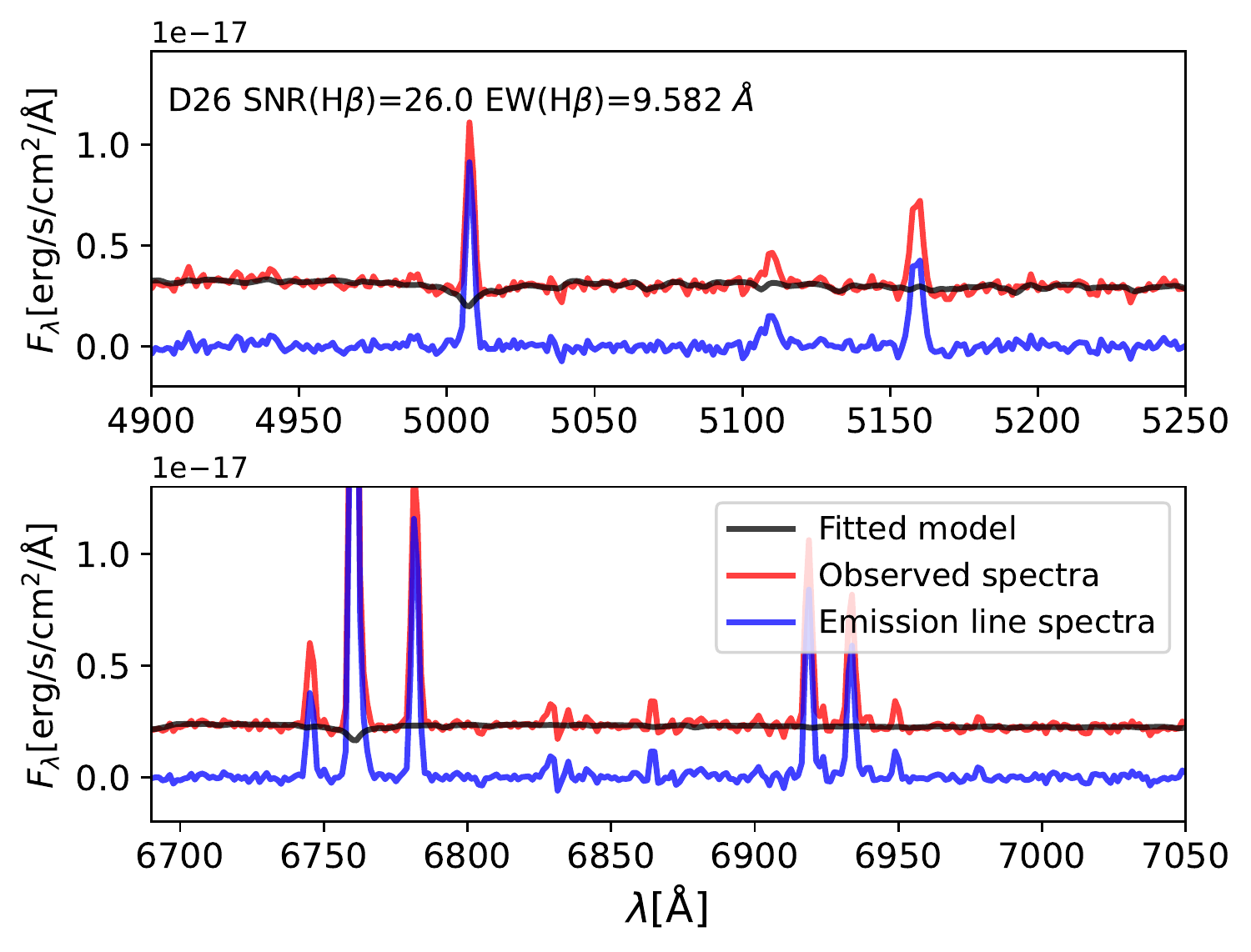}\\
\includegraphics[width=0.5\linewidth]{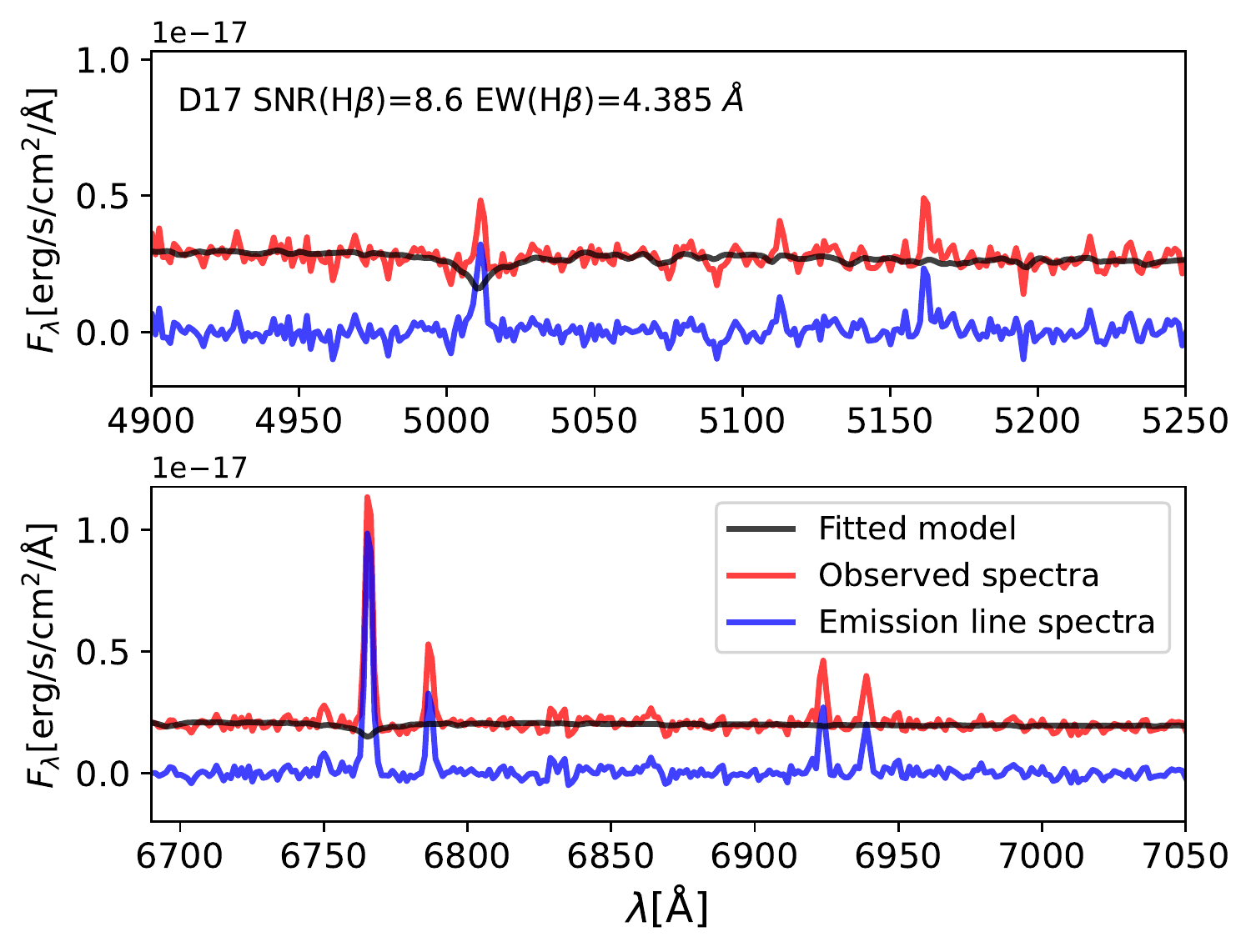}\\
\par\end{centering}
\caption{
 Subtraction of the underlying stellar continuum; 
spectra of three regions from the inner disk: D40 (top left), D26 (top right), and D17 (bottom), are shown in two sub-panels each, the top one covering the \hb\ line, and the bottom one, the \ha\ line. In each plot, the observed, fitted model and the pure emission line spectra are shown in red, grey and blue, respectively. The three plotted spectra illustrate the effect of the underlying absorption on the measured \ha\ and \hb\ lines for relatively high, medium and low measured \hb\ SNRs and \hb\ EWs, respectively. The observed, not the rest-frame, spectra are shown in all cases. 
}
\label{ssp_fit}
\end{figure*}

We subtracted the fitted continuum to obtain pure nebular spectra, which is the one we used for the abundance analysis  in the inner disk regions. 
We illustrate the fitting and subtraction of the underlying absorption feature in Fig.~\ref{ssp_fit} for three cases: the top-left, top-right and bottom panels show spectra for inner disk regions with relatively high, intermediate and low \ewhb; these three cases also correspond to high, intermediate and low SNR(\hb), respectively. The model spectra (black line) fit very well the observed spectra (blue line) in all three illustrated cases, and hence the model-subtracted spectrum corresponds to a pure nebular spectrum. As expected, the nebular \hb\ line is progressively more affected by the underlying absorption as the nebular \ewhb\ decreases.

 Unlike the disk \hii\ region spectra, the ring \hii\ region spectra are dominated by nebular lines with no strong-enough absorption feature for a reliable fitting of the continuum with an underlying stellar population. Nevertheless, we took care of the possible presence of weak absorption features at \ha\ and \hb\ wavelengths by adding 2~\AA\ to the EWs of these two lines following the standard procedure advocated by \citet{McCall1985}.

\subsection{Measurement of line fluxes}

The main purpose of the spectral analysis is to measure accurate fluxes of emission lines suitable for nebular abundance measurements. The spectral range of MUSE includes
temperature-sensitive lines of \nii\ and \siii\ ions, density-sensitive lines of
\sii, \cliii\ and \ariv\ ions, in addition to lines for determination of abundances of He, O, N, S, Ar, Cl and Fe. 
The complete list of lines we measured is available as a supplementary material in a machine readable table (MRT) format.
We used the Gaussian profile fitting routine of the {\it splot} task in {\sc iraf}
for measuring the fluxes of individual spectral lines. We fitted Gaussian profiles at the redshifted wavelength of each line in the list in an  automated way using the cursor command facilities of IRAF. The flux, equivalent width (EW), full width at half maximum (FWHM) of each fitted line are stored for analysing the signal-to-noise ratio (SNR=flux/$\sigma_{\rm l}$) of each line.
The noise in measured flux ($\sigma_{\rm l}$) of each line is calculated using
the expression \citep{Tresse1999}: \\

\begin{equation}
\sigma_{\rm l} = \sigma_{\rm c} D \sqrt{(2 N_{\rm pix} + \frac{EW}{D})},
\label{eqn:noise}
\end{equation}
where $D$ is the spectral dispersion in \AA\ per pixel (1.25 for MUSE), $\sigma_{\rm c}$ is
the mean standard deviation per pixel of the continuum, which is measured in a
line-free part of the continuum adjacent to each line, $N_{\rm pix}$ is the
number of pixels covered by the line, which is equated to the FWHM of the fitted Gaussian profile. We imposed the condition that for a line to be considered detected, it should
have a SNR$\ge$4 and FWHM 
comparable to that of the \hb.
We used a factor of 2 tolerance in FWHM, which takes into account errors in
the FWHM of the fitted Gaussians, especially for the lines at the threshold
of detection (flux $\lesssim$1\% of \hb), such as 
the temperature sensitive lines of \niia\ and \siiia. When a line does
not fulfill these criteria, it is assigned a limiting flux of 3$\times\sigma_{\rm l}$, where we substituted $N_{\rm pix}$ by the FWHM of the \hb\ line.

The \ha-emitting regions detected in the region between the nucleus and the outer ring are, in general, fainter in all lines as compared to their outer ring counterparts.
Temperature-sensitive lines are not traced in  any of these regions. 
 The diffuse ionized gas (DIG) component can affect the line measurements. Since DIG is related to young and old stars \citep{2022A&A...659A..26B}, it is not clear if DIG removal is better for studies like the one presented here. However, we have estimated how much our line estimates can be affected by the DIG. We have defined diffuse regions of the ring with H$\alpha$ surface brightness $F_{\rm{H\alpha}}<5\times 10^{-17}\rm{erg/s/cm^2/arcsec^2}$, and extracted the spectrum of these regions. After subtracting the DIG spectrum from the HII regions, we found that the difference in [NII], [SII], [OIII]  line fluxes are of the order of 0.01dex for the brightest regions, while as much as 0.08dex for the faintest regions. Therefore, we decide not remove the DIG since its contribution does not affect the results presented in this work.

The \ha\ and \hb\ line fluxes for all the regions where the \hb\ line is detected at SNR$\ge$4 are used to determine the attenuation from the dust in the \hii\ region and along the line of sight. For this purpose, we used the Balmer decrement method for case B recombination of a typical photoionized nebula \citep{Osterbrock2006} and the reddening curve of \citet{Cardelli1989}.
 The Balmer recombination line ratios are only weakly dependent on $T_{\rm e}$ and $n_{\rm e}$ values, and hence we used an intrinsic $\frac{I(\rm H\alpha)}{I(\rm H\beta)}=2.863$ corresponding to $T_{\rm e}$=10000~K, $n_{\rm e}$=100 cm$^{-3}$, canonical values for extragalactic \hii\ regions. The choice of these canonical values for the Cartwheel is justified by the directly determined $T_{\rm e}$ values for the ring \hii\ regions (11000--15000~K) and those estimated for the disk regions ($\sim$8000-10000~K) based on the empirical relations of \citet{2016MNRAS.457.3678P} between the O abundance and $T_{\rm e}$. The densities in the Cartwheel \hii\ regions are also close to the canonical value used. The \Av\ values would differ from the calculated values at the most by $\pm$0.04~mag using the measured/estimated values for each region instead of the canonical values, which is small compared to the range of measured \Av\ values.

The attenuation-corrected line fluxes normalized to that of the \hb\ line are tabulated in the supplementary MRT table. The error on the line fluxes normalized to that in \hb\ is calculated by propagating the errors on the line of interest and the \hb\ line. The table also contains the \Av\ values along with the determined errors, the corrected \hb\ flux and the SNR,  \ha\ and \hb\ EWs, as well as the coordinates. For spectra having SNR(\hb)$\ge40$, we tabulate the 1-$\sigma$ upper limits of all lines in the error column.

\section{Determination of abundances}
\subsection{Determination of ionic and elemental abundances from direct method}

Surprisingly, the last published work on the measurement of metallic abundances
in the Cartwheel is by \citet{Fosbury1977}. Using spectra taken with photographic
plates, they reported He, O, N and Ne abundances for two of the brightest regions,
referred to as knots A and B, which correspond to our \hii\
region numbers \#99 and \#119, respectively. The average O abundance
they obtained for these two knots is 12+$\log$(O/H)=8.0$\pm$0.1. Using Balmer
decrement method for a photoionized nebula, they reported an $E(B-V)=0.64$,
which corresponds to \Av=2~mag for the Galactic extinction curve with \Rv=3.1.

\begin{figure}
\begin{centering}
\includegraphics[width=1.0\linewidth]{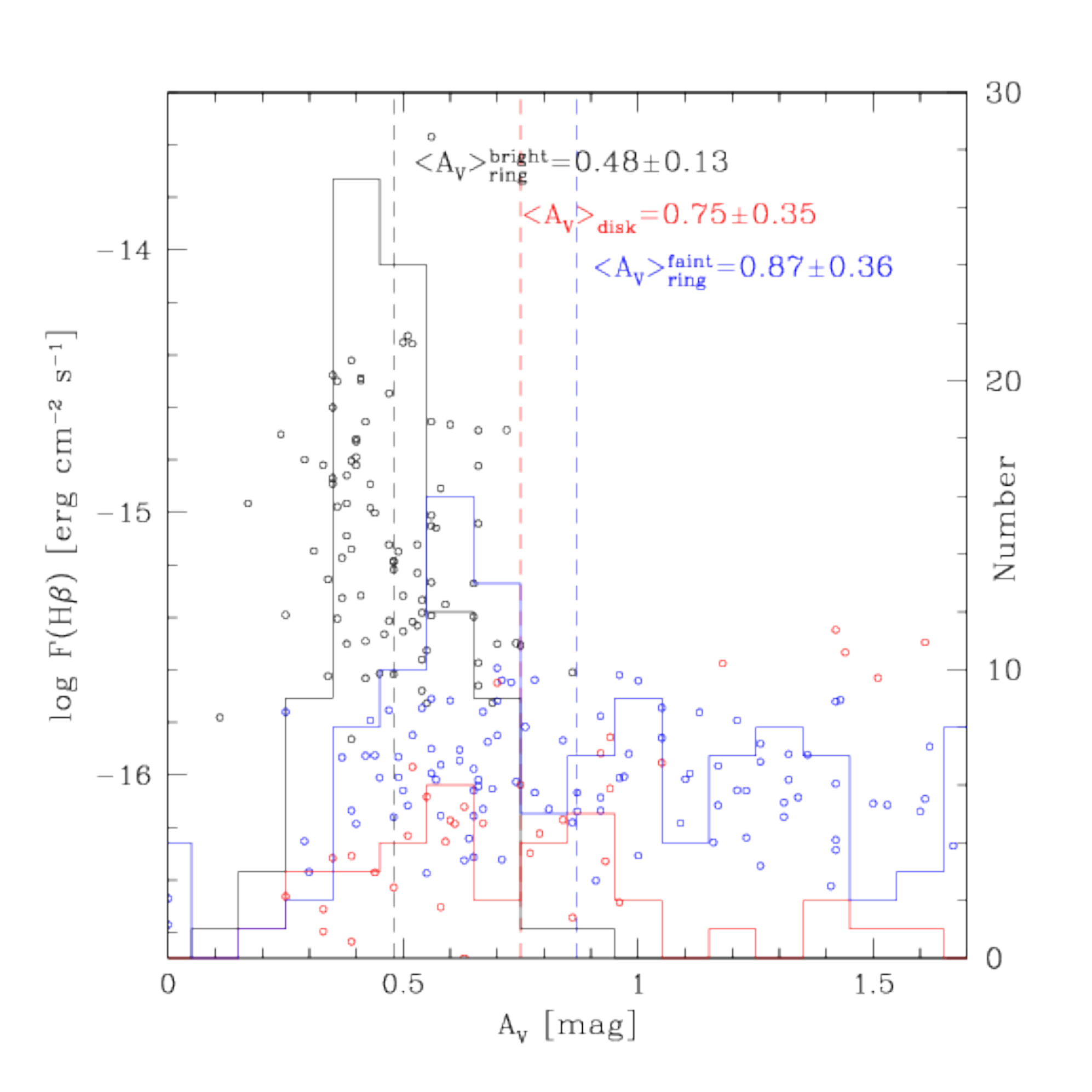}
\par\end{centering}
\caption{Distribution of \Av\ for \hii\ regions in the Cartwheel (histogram
 at the right axis scale). The points show the distribution of the reddenning
corrected \hb\ flux in each bin of \Av. The values for ring and disk \hii\ regions are shown in blue and red colours, respectively. The average values of \Av\ for ring and disk samples are shown by the vertical dashed lines of blue and red colours, respectively. The disk \hii\ regions are systematically fainter and reddened than the ring \hii\ regions.
}
\label{fig:heii_av}
\end{figure}

The MUSE dataset has allowed us to obtain \Av\ values using \hii\ regions in the ring and the disk. In Fig.~\ref{fig:heii_av}, we show the distribution of these \Av\ values, divided into three different groups: bright ring regions, consisting of 88 regions that have at least a SNR of 40
in the \hb\ line; the remaining 103 relatively fainter ring \hii\ regions; and 36 of the 40 disk \hii\ regions that had detectable \hb\ emission. 

The mean value for the bright ring sample is 0.48$\pm$0.13~mag,  with the 
\Av\ values for the two regions in common with \citet{Fosbury1977} are 0.56~mag (\#99) and 0.35~mag (\#119).  These values are significantly smaller than the often used value of \Av=2~mag, based on the photographic plate spectra of \citet{Fosbury1977}. We believe the non-linearity in the flux calibration of photographic spectra is the reason for the high \Av\ values reported by \citet{Fosbury1977}.
 In comparison, the fainter ring \hii\ regions and the disk \hii\ regions have mean values of 0.87$\pm$0.36~mag and 0.75$\pm$0.35~mag, respectively, which are systematically higher as compared to the bright ring regions. 
We use the measured \Av\ to deredden all the line fluxes.

 The range of \Av\ values in the Cartwheel are in the range of observed values in extragalactic \hii\ regions \citep[e.g.][]{2015A&A...574A..47S,2020MNRAS.494.1622E}.
However, there is an important difference between the Cartwheel and normal galaxies: brighter regions have systematically higher extinction in normal galaxies, whereas Fig.~\ref{fig:heii_av} shows a clear tendency for the brighter \hii\ regions to have lower \Av\ values. The direct correlation between \hb\ flux and \Av\ is expected for regions that follow the Kennicutt-Schmidt relation \citep{2020MNRAS.492.2651B} between the SFR and gas density, as the \hb\ flux is a direct tracer of SFR and \Av\ is proportional to the gas density. Conversely, the anti-correlation between \hb\ flux and \Av\ in the Cartwheel indicates a breakdown of Kennicutt-Schmidt relation. Such a breakdown in the Cartwheel has been comprehensively illustrated by \citet{Higdon2015}, who obtained total gas densities in the ring regions by combining recent ALMA 
observations of the molecular gas and the \citet{Higdon1996} \hi\ observations. The most likely reason for the breakdown is the negative feedback, namely destruction of molecular clouds and expulsion of gas, due to winds from massive stars and supernovae (SNe) explosions associated to the recent star formation.

\begin{figure*}
\begin{centering}
\includegraphics[width=0.495\linewidth]{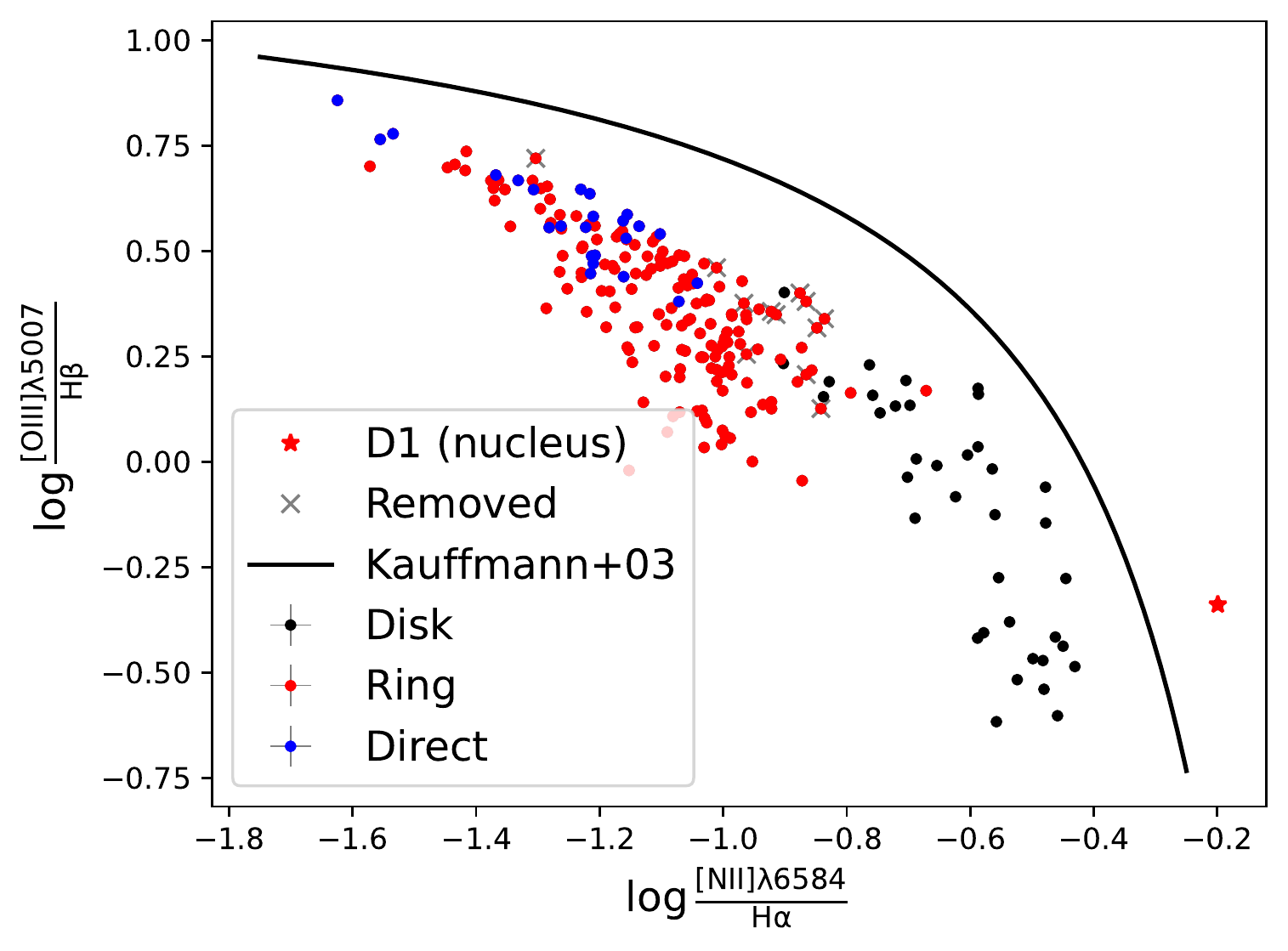}~
\includegraphics[width=0.495\linewidth]{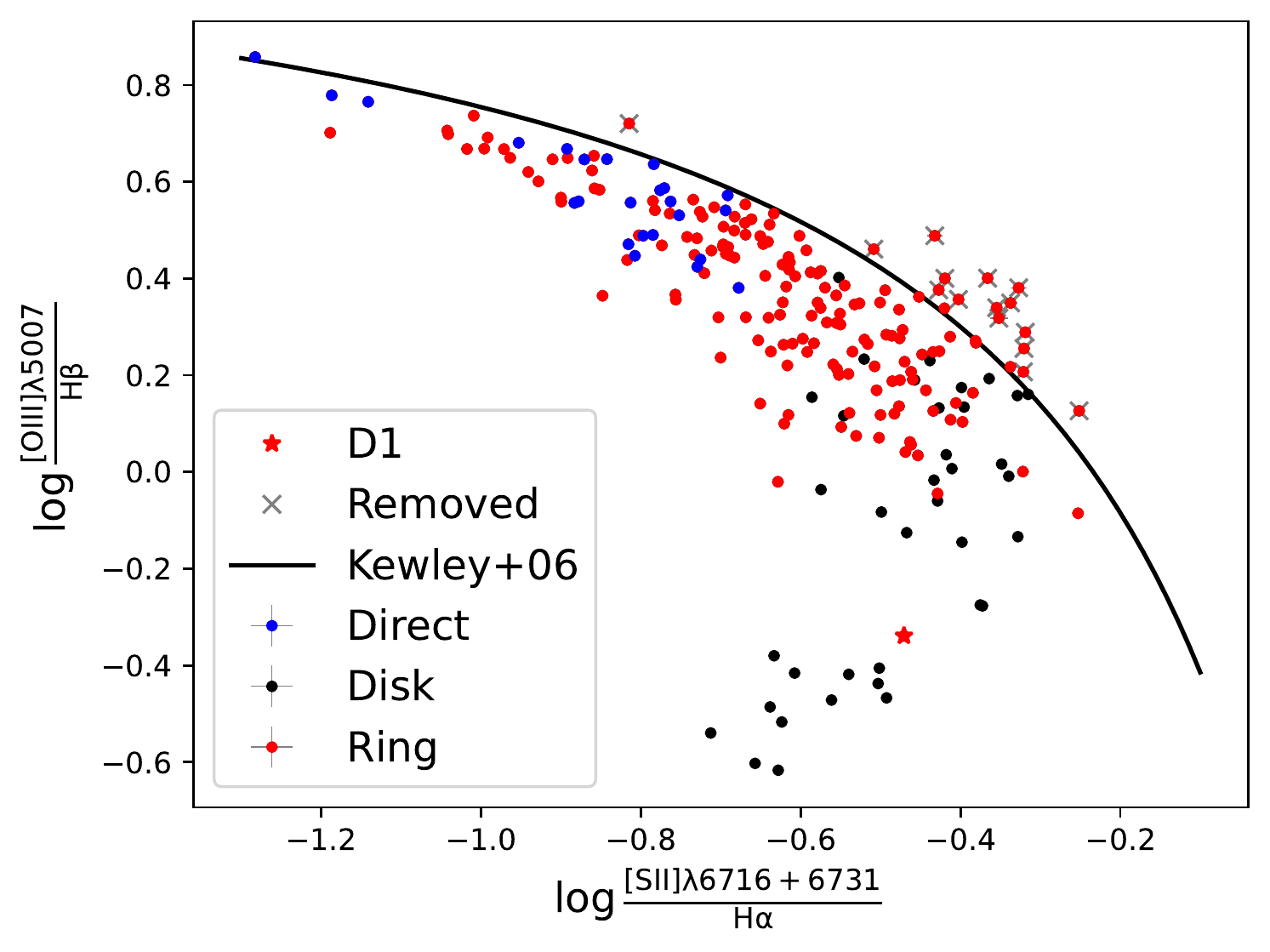}\\
\par\end{centering}
\caption{ BPT diagrams for the identified regions in Cartwheel; (left)  $\rm{\log\frac{[OIII]\lambda5007}{H\beta}}$ vs $\rm{\log\frac{[NII]\lambda6584}{H\alpha}}$ and (right) $\rm{\log\frac{[OIII]\lambda5007}{H\beta}}$ vs $\rm{\log\frac{[SII]\lambda6717+31}{H\alpha}}$.
We show the regions from the external ring (red dots), from the inner disk (black dots), the regions where DM abundances were determined (blue dots) and the nuclear region (D1) (red star).
Regions photoionized by stars are expected to lie below the curves defined by \citet{2003MNRAS.346.1055K} (left) and \citet{2006MNRAS.372..961K} (right). Regions marked by a grey cross may have contribution from shock ionization, and were omitted for the analysis of abundances.
}
\label{fig_bpt}
\end{figure*}

 In order to ensure that the emission lines we have measured are due to photoionization associated to star formation, we show in Fig. \ref{fig_bpt} 
both the ring and the inner disk regions in 
the BPT diagrams \citep{1981PASP...93....5B}. 
We show as a black line the curve used to differentiate between star formation and AGNs from \citet{2003MNRAS.346.1055K,2006MNRAS.372..961K}. In the case of the use of [NII] (Fig. \ref{fig_bpt} left), it is clear that all regions, except one, are well inside the boundary defined for the photoionization by stars. The exception is the region labelled as
D1 in Fig. \ref{fig:muse_image} which corresponds to the nucleus of the Cartwheel,  
and shows LINER-like spectral characteristics. We exclude this region from the analysis in the rest of the paper. In the case of the use of [SII] (Fig. \ref{fig_bpt} right) there are a few regions clearly outside the empirical division for photoionization \citep{2006MNRAS.372..961K}. We do not use those regions marked as rejected in the following analysis. 
Thus, all the regions that are analysed for the determination of nebular abundances are associated to star formation.

Chemical abundances were calculated using the ``direct method'' (DM), which requires
the measurement of the physical condition of the ionized gas, namely the
electron temperature ($T_\mathrm{e}$) and the electron density ($N_\mathrm{e}$).
We used the three zone (high, medium and low) ionization model of \hii\ regions proposed by
\citet{1992AJ....103.1330G} to determine the ionic abundances.
The bluest wavelength of MUSE spectral coverage is $\sim$4620~\AA\ at the rest
wavelength for Cartwheel, which implies that \oiiia, the most
commonly used temperature-sensitive line, cannot be used for the determination of
the temperature of the high ionization zone ($T_{\rm{e}}^{\rm{high}}\equiv T_\mathrm{e}$(\oiii)).
However, the spectral range covers the \siiib\ line at the red border for majority of the regions which, when combined with \siiia\ line, measures the temperature of the medium ionization zone ($T_{\rm{e}}^{\rm{med}}\equiv T_\mathrm{e}$(\siii)). The regions without the \siiib\ line fluxes are the ones in the western part of the ring (identification numbers between 22 and 96), which have recessional velocities that shift the line outside the spectral coverage of MUSE. Both the \siiia\ and \siiib\ lines are detected above an SNR$>$4 in 29 of our regions.
The spectral range covers all the three \nii\ lines for measuring the
$T_\mathrm{e}$(\nii), the temperature of the low ionization zone ($T_{\rm{e}}^{\rm{low}}$). 
We notice that in the Cartwheel the \niia\ is systematically fainter than the \siiia, allowing the direct determination of $T_{\rm{e}}^{\rm{low}}$ in only  5 of the 24 regions where $T_{\rm{e}}^{\rm{med}}$ has been determined.  For the sake of uniformity in the temperature estimations, we preferred to estimate $T_{\rm{e}}^{\rm{low}}$  from $T_{\rm{e}}^{\rm{med}}$ as explained below for all regions, even for the five regions where the \niia\ is detected. 

We used the equations obtained by \citet{1992AJ....103.1330G} using photoionization models to relate the temperatures in the three zones. The $T_\mathrm{e}$ for the high ionization zone was obtained for all the regions with $T_{\rm{e}}^{\rm{med}}$ measurements using the relation:
 \begin{equation}
  T_{\rm{e}}^{\rm{high}}=1.2 \thinspace T_{\rm{e}}^{\rm{med}}-2048\thinspace \rm{K},
  \label{eqtemp1}
 \end{equation}
and $T_{\rm{e}}^{\rm{low}}$
using:
\begin{equation}
  T_{\rm{e}}^{\rm{low}}=0.7\thinspace  T_{\rm{e}}^{\rm{high}}+3000\thinspace \rm{K}. 
  \label{eqtemp2}
 \end{equation}
Both relations are supported by observational data \citep[e.g.,][]{2003ApJ...591..801K,2009ApJ...700..654E,2009ApJ...700..309B, Karla2020}.

The wavelength range covered by the MUSE spectra covers the density-sensitive lines of \sii, \cliii\  and \ariv, which originate in low, medium and high ionization zones, respectively. However, the doublets of \cliii\ and \ariv\ are faint to obtain reliable measurement of $N_\mathrm{e}$ corresponding to the zones in which these ions are found. On the other hand, both the \sii\ lines are detected in all the  24 regions, where the temperature-sensitive \siii\ line is detected. We hence obtained density from \sii\ lines and assumed it to be uniform over the entire \hii\ region for the sake of obtaining abundances from DM.

We used PyNeb \citep{2015A&A...573A..42L} to determine simultaneously the $T_\mathrm{e}$(\siii) and $N_\mathrm{e}$(\sii) using \siii$\lambda6312$/$\lambda9069$ and  \sii$\lambda6717$/$\lambda6731$, respectively, for the 24 regions where both the \siiia\ and \siiib\ lines have been detected. 
The relative flux ratio uncertainties are propagated as the relative uncertainties of $T_\mathrm{e}$ and $N_\mathrm{e}$.

\begin{table*}
\small\addtolength{\tabcolsep}{-2pt}
\caption{$T_\mathrm{e}$, $N_\mathrm{e}$ and element abundances determined with the DM for some of the regions in Cartwheel.  \label{tab_abundances_direct}}
\centering
\begin{tabular}{lcccccccc}
ID & $T_{\rm{e}}^{\rm{low}}\thinspace^{\rm{(a)}}$ & $T_{\rm{e}}^{\rm{high}}\thinspace^{\rm{(a)}}$ & $T_{\rm{e}}^{\rm{medium}}$ & $N_\mathrm{e}$(\sii) & 12+log(O/H) & 12+log(Fe/H) & 12+log(N/H) & 12+log(He/H) \\
 & $\times10^{4}$ K  & $\times10^{4}$ K  & $\times10^{4}$ K  & $\rm{e^{-}/cm^3}$ &  &  &  & \\
 \hline
9   & 1.11 $\pm$ 0.04 & 1.15 $\pm$ 0.05 & 1.05 $\pm$ 0.04 &  50 $\pm$ 5  & 8.38$\pm$0.11 & 5.88$\pm$0.15 & 6.68$\pm$0.10 & \dots \\
10  & 1.12 $\pm$ 0.05 & 1.17 $\pm$ 0.08 & 1.07 $\pm$ 0.06 & 27.9$\pm$3.3 & 8.35$\pm$0.14 & 5.93$\pm$0.17 & 6.71$\pm$0.12 & \dots \\
12  & 1.32 $\pm$ 0.18 & 1.45 $\pm$ 0.25 & 1.41 $\pm$ 0.21 &  55 $\pm$ 22 & 8.0 $\pm$0.4  & \dots            & 6.5 $\pm$0.4  & \dots \\
14  & 1.20 $\pm$ 0.12 & 1.28 $\pm$ 0.18 & 1.20 $\pm$ 0.15 &  58 $\pm$ 13 & 8.16$\pm$0.28 & \dots            & 6.60$\pm$0.24 & \dots \\
16  & 1.21 $\pm$ 0.05 & 1.29 $\pm$ 0.07 & 1.22 $\pm$ 0.05 & 55.1$\pm$3.5 & 8.19$\pm$0.08 & 5.87$\pm$0.09 & 6.62$\pm$0.07 & 10.93$\pm$0.11 \\
60  & 1.28 $\pm$ 0.05 & 1.40 $\pm$ 0.07 & 1.34 $\pm$ 0.06 & 109 $\pm$ 29 & 8.07$\pm$0.28 & 5.7 $\pm$0.4  & 6.39$\pm$0.26 & 10.89$\pm$0.31 \\
98  & 1.136$\pm$ 0.033& 1.19 $\pm$ 0.05 & 1.10 $\pm$ 0.04 & 29.2$\pm$2.6 & 8.34$\pm$0.10 & 5.95$\pm$0.12 & 6.63$\pm$0.09 & \dots \\
99  & 1.250$\pm$ 0.006& 1.357$\pm$ 0.008& 1.295$\pm$0.007 & 185 $\pm$ 7  & 8.16$\pm$0.04 & 5.92$\pm$0.07 & 6.44$\pm$0.04 & 10.94$\pm$0.04 \\
100 &  1.16$\pm$ 0.04 & 1.23 $\pm$ 0.05 & 1.14 $\pm$ 0.04 & 27.9$\pm$1.2 & 8.30$\pm$0.06 & 6.01$\pm$0.10 & 6.61$\pm$0.05 & 10.92$\pm$0.07 \\
103 &  1.17$\pm$ 0.06 & 1.25 $\pm$ 0.09 & 1.17 $\pm$ 0.07 & 25.6$\pm$1.4 & 8.23$\pm$0.09 & 5.96$\pm$0.11 & 6.61$\pm$0.08 & \dots \\
105 &  1.20$\pm$ 0.04 & 1.28 $\pm$ 0.06 & 1.21 $\pm$ 0.05 &  18 $\pm$ 6  & 8.20$\pm$0.33 & 5.87$\pm$0.33 & 6.53$\pm$0.31 & \dots \\
106 & 1.191$\pm$ 0.030& 1.27 $\pm$ 0.04 & 1.195$\pm$ 0.035&  24 $\pm$ 5  & 8.21$\pm$0.21 & 5.82$\pm$0.25 & 6.51$\pm$0.19 & 10.90$\pm$0.24 \\
108 &  1.21$\pm$ 0.11 & 1.30 $\pm$ 0.16 & 1.23 $\pm$ 0.13 & 18.8$\pm$2.2 & 8.11$\pm$0.17 & 5.95$\pm$0.19 & 6.56$\pm$0.15 & 10.89$\pm$0.17 \\
111 &  1.21$\pm$ 0.06 & 1.30 $\pm$ 0.08 & 1.22 $\pm$ 0.07 &  44 $\pm$ 7  & 8.19$\pm$0.17 & 5.96$\pm$0.17 & 6.51$\pm$0.15 & \dots \\
112 &  1.53$\pm$ 0.05 & 1.76 $\pm$ 0.08 & 1.77 $\pm$ 0.06 &  13 $\pm$ 6  & 7.8 $\pm$0.5  & 5.5 $\pm$0.7  & 6.4 $\pm$0.4  & \dots \\
119 &  1.33$\pm$ 0.07 & 1.47 $\pm$ 0.10 & 1.43 $\pm$ 0.08 & 1.9 $\pm$1.3 & 8.0 $\pm$0.7  & 5.8 $\pm$0.7  & 6.5 $\pm$0.7  &  10.9$\pm$0.7 \\
120 &  1.30$\pm$ 0.10 & 1.43 $\pm$ 0.14 & 1.38 $\pm$0.12  & 12.7$\pm$3.4 & 8.02$\pm$0.29 & 5.83$\pm$0.32 & 6.46$\pm$0.27 & 10.90$\pm$0.32 \\
122$^{\rm{(b)}}$&1.55$\pm$0.23&1.78$\pm$0.33&1.80$\pm$0.28&   48$\pm$14  & 7.8 $\pm$0.4  & \dots            & 6.44$\pm$0.33 & \dots \\
126$^{\rm{(b)}}$&1.8 $\pm$0.4 &2.2 $\pm$0.5 & 2.2$\pm$0.4 &   80$\pm$40  & 7.5 $\pm$0.6  & \dots            & 6.2 $\pm$0.6  & \dots \\
129$^{\rm{(b)}}$&1.53$\pm$0.29&1.8 $\pm$0.4 &1.77$\pm$0.35&   48$\pm$11  & 7.74$\pm$0.32 & \dots            & 6.35$\pm$0.29 & \dots \\
131$^{\rm{(b)}}$&1.55$\pm$0.27&1.8 $\pm$0.4 &1.81$\pm$0.32&   36$\pm$18  & 7.7 $\pm$0.6  & \dots            & 6.4 $\pm$0.5  & \dots \\
141 &  1.14$\pm$ 0.06 & 1.20 $\pm$ 0.08 & 1.11 $\pm$ 0.07 & 57.5$\pm$3.2 & 8.31$\pm$0.08 & 5.99$\pm$0.09 & 6.58$\pm$0.08 & \dots \\
144 &  1.31$\pm$ 0.09 & 1.45 $\pm$ 0.13 & 1.40 $\pm$0.11  &  72 $\pm$13  & 8.00$\pm$0.20 & 5.79$\pm$0.20 & 6.47$\pm$0.19 & 10.92$\pm$0.20 \\
156 &  1.18$\pm$ 0.14 & 1.25 $\pm$ 0.20 & 1.17 $\pm$0.16  &  60 $\pm$10  & 8.17$\pm$0.23 & \dots            & 6.69$\pm$0.20 & \dots \\
\hline
\end{tabular}
\\
{ $^{\rm{a}}$ Not directly estimated from observations, but using Eqs. \ref{eqtemp1} and \ref{eqtemp2}. 
 $^{\rm{b}}$ Rejected for the DM analysis due to its high determined temperature as compared to that expected from the T$_{\rm{e}}-N_{2\rm{H\beta}}$ relation \citep{2016MNRAS.457.3678P}.
}   \\
\hrulefill 
\label{tab_direct}
\end{table*}

Ionic abundances are calculated using PyNeb, where we use as inputs the
$T_{\rm{e}}$ corresponding to each ionization zone, the $N_{\rm{e}}$ estimated
previously, as well as the following lines (when detected) for each ion:
$\rm{Fe^{2+}}$~$\lambda\lambda$4658, 4881, 4986, 5270;
$\rm{O^{0}}$~$\lambda$6300;
$\rm{O^{+}}$~$\lambda\lambda$7320, 7330;
$\rm{O^{2+}}$~$\lambda\lambda$4959, 5007;
$\rm{N^{+}}$~$\lambda\lambda$6549, 6583;
 $\rm{He^{+}}$~$\lambda\lambda$4922, 5876, 6678, 7065, 7281; and 
$\rm{He^{++}}$~$\lambda$4686.
The relative uncertainty of the ionic abundance for each line is determined from
the quadratic sum of the relative $T_{\rm{e}}$, $N_{\rm{e}}$ and flux ratio
uncertainties. The ionic abundances are obtained by an error weighted average
of the ionic abundance for each line.

Finally, we determined the abundance for each element. For O and He we used the
direct sum of their ionic abundances in their three states:
N(O)=N(O$^0$)+N(O$^+$)+N(O$^{2+}$) and N(He) =  N(He$^+$)+N(He$^{++}$).
 The equation for He abundance assumes that there is no neutral He in the H ionized volume, which is valid within 2\% for high ionization regions that show the \heiiwr\ line \citep{2013A&A...558A..57I}. 
The \heiiwr\ line is detected in all the 20 regions that are used for abundance measurement from DM (see Mayya et al. submitted to MNRAS). We report here the He abundance only for the regions where the SNR of the \heiiwr\ is larger than 4, which  is fulfilled for 9 regions. 
The presence of He$^{+}$ ion implies that some fraction of O could
be in O$^{3+}$ state.  We used the ionization correction factor which considers the O$^{3+}$ abundance for the high excited HII regions from  \citet{2006A&A...448..955I}, although the change in the total O abundance because of this correction is as small as 0.7\%.
For the Fe and N abundances we used the $\rm{O^{+}}$, $\rm{O^{2+}}$,
with the corresponding element ionic abundances and the intermediate metallicity
($7.6\leq 12+\log\frac{\rm{O}}{\rm{H}}\leq 8.2$) ionization correction factors from
\citet{2006A&A...448..955I}. We present the $T_\mathrm{e}$, $N_\mathrm{e}$ and element abundances used in this work in Tab. \ref{tab_abundances_direct}.

\subsection{Determination of elemental abundances from strong-line method (SLM)}

The abundances estimated using the DM  are viable just for  24
regions in the external ring. In order to estimate the O abundances over the whole system,  we used empirical strong line calibrations. In particular, we choose the 
``S calibrator'' from  \citet{2016MNRAS.457.3678P} as the fiducial one for this work, to determine the O abundances of all nebular regions.
The ``S calibrator'' is chosen because it is the only calibrator in the MUSE spectral range that estimates metallicity taking into account the dependence of line ratios on both $T_{\rm e}$ and ionization parameter. The ``S calibrator'' 
uses the intensities of the following strong lines: \hb, \ha,  \oiii$\lambda$4959+$\lambda$5007,  \nii$\lambda$6548+$\lambda$6584 and \sii$\lambda$6717+$\lambda$6731.

The \oiii$\lambda$4959 line for some of the regions, especially in the disk, is only marginally detected. However, the flux of this line is  theoretically one third of the \oiii$\lambda$5007 line flux \citep[see e.g.,][]{Perez-Montero2017}. Hence, for the sake of consistency, for all regions we estimated the \oiii$\lambda$4959 line flux from the \oiii$\lambda$5007, using this constant factor.
We have derived the O abundance with this method for 152 regions in the external ring, and  for 36 regions within the disk. Our results are included in the supplementary MRT table. 

\begin{figure*}
\begin{centering}
\includegraphics[width=0.495\linewidth]{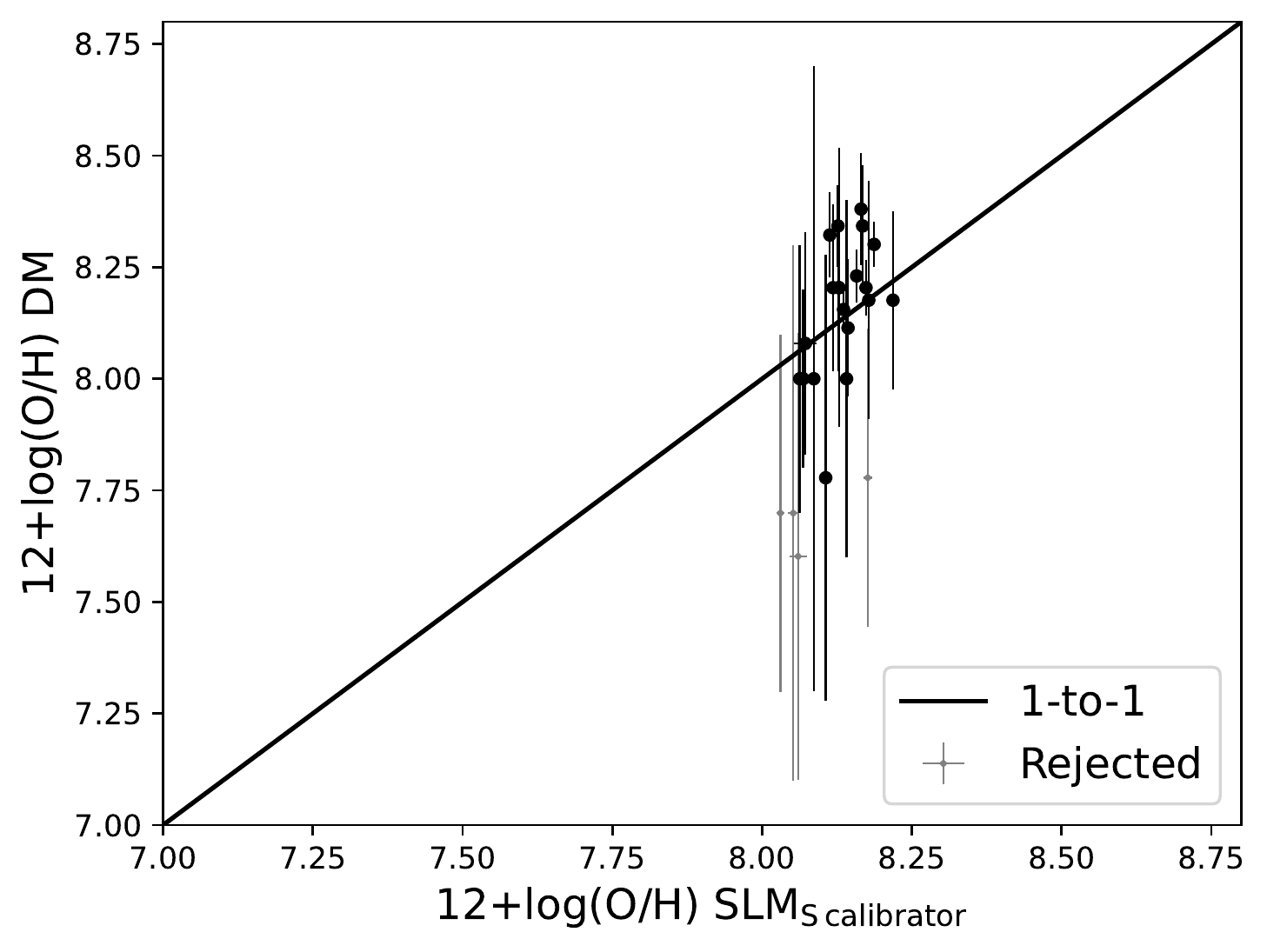}~
\includegraphics[width=0.495\linewidth]{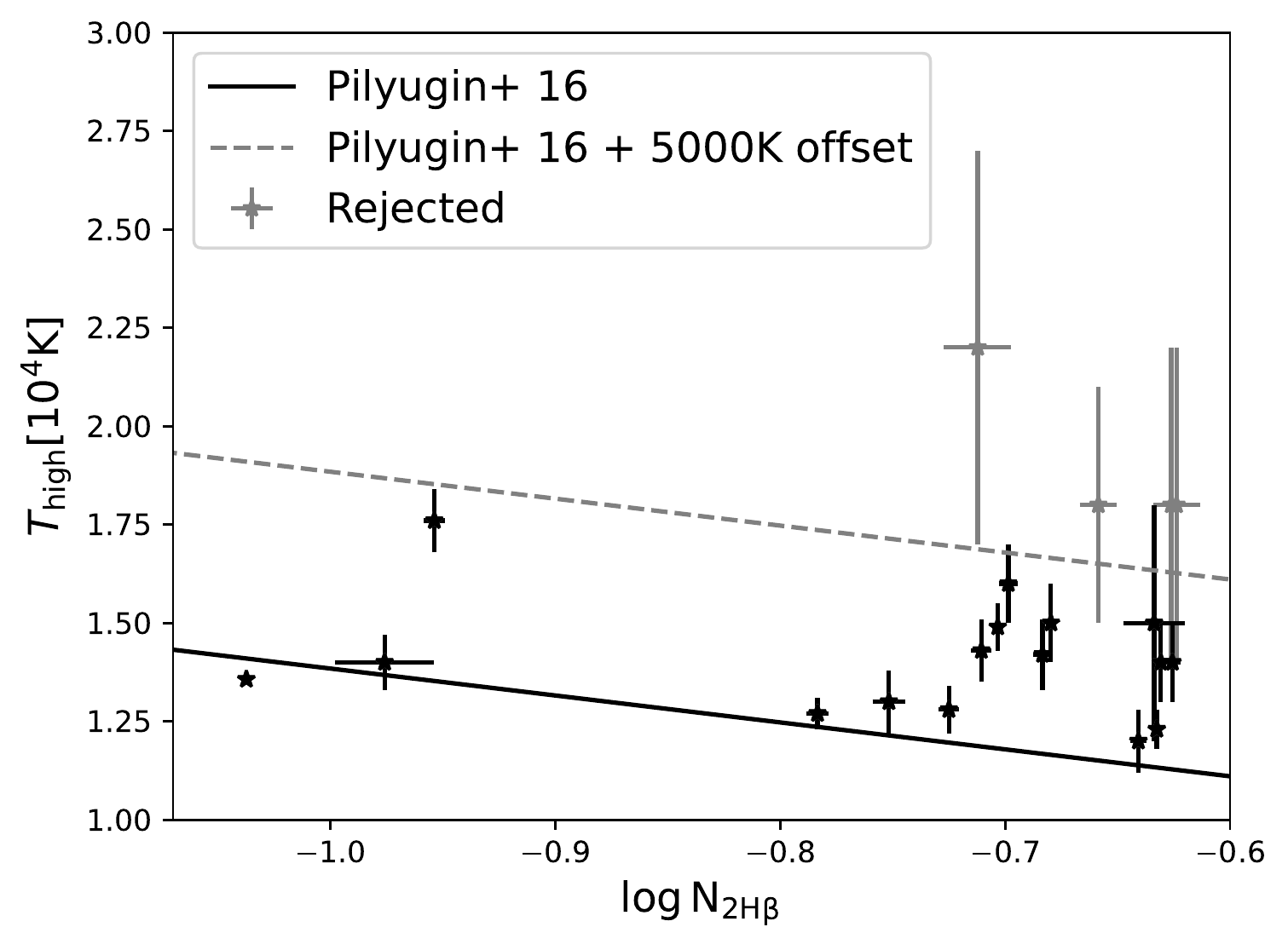}\\
\par\end{centering}
\caption{ (left) O abundance with the DM versus that obtained with the  S calibrator from \citet{2016MNRAS.457.3678P}; (right) Temperature of the high ionization zone, $T_{\rm{e}}^{\rm{high}}$, versus $\log\rm{N_{2H\beta}}$ index, $\rm{\frac{[NII]\lambda6548+\lambda6584}{H\beta}}$. The solid line is the tight relation reported in \citet{2016MNRAS.457.3678P}; the grey dashed line is the same relation shifted upwards by 5000~K. Estimated $T_{\rm{e}}^{\rm{high}}$ for regions above the dashed line are unlikely to be reliable and hence are rejected (grey points) from the abundance analysis using the DM.
}
\label{fig:direct_strong}
\end{figure*}

In Fig. \ref{fig:direct_strong} (left) we compare the  O abundances obtained using the DM and  the S calibrator for the \hii\ regions in the outer ring. 
The unit-slope line passes through the majority of the points within the calculated errors, implying a general agreement in the abundances determined from the two methods. However, the $12+\log\frac{\rm{O}}{\rm{H}}$ values obtained from DM of 5 regions (shown in grey) are below 8.0, the minimum value obtained from SLM. The inferred low metallicities could be due to overestimation of temperature. In order to investigate this, we plot in Fig.~\ref{fig:direct_strong} (right) the temperature of the high ionization zone, $T_{\rm{e}}^{\rm{high}}$, versus $\rm{N_{2H\beta}}$ index ($\rm{N_{2H\beta}}=\rm{\frac{[NII]\lambda6548+\lambda6584}{H\beta}}$).
\citet{2016MNRAS.457.3678P} have found the $\rm{N_{2H\beta}}$ ratio to be a good indicator of temperature. We plot their relation in this plot by a solid line. We also indicate in this plot a line that corresponds to an upward offset of 5000~K as compared to the \citet{2016MNRAS.457.3678P} relation. The same regions for which the inferred $12+\log\frac{\rm{O}}{\rm{H}}<8.0$ also have the measured temperatures that are higher by as much as 5000~K, as compared to the expected relation.

Recently, \citet{Menacho2021} reported similar O abundance discrepancies between the DM and SLM in some regions of the interacting galaxy Haro 11. They argued that the presence of shocks can overestimate the temperature determination due to a non Maxwellian distribution of electron velocities \citep{1980ARA&A..18..219M}, or due to the increase of the flux of auroral lines used for the temperature determination \citep{1991PASP..103..815P}. 
As illustrated in Fig.~\ref{fig_bpt}, there is no direct evidence for the presence of shocks in the Cartwheel \hii\ regions, and hence the origin of this discrepancy possibly requires a different explanation.

\citet{Karla2020} have demonstrated that the O abundances determined from DM tend to be underestimated by more than 0.2 dex for high excitation regions, when using $T_{\rm e}$ relations to obtain $T_{\rm{e}}^{\rm{high}}$ rather than obtaining it directly from \oiiia\ line. This is due to the large dispersion in the relations given by equations \ref{eqtemp1} and \ref{eqtemp2}.
Unfortunately, the accuracy of O abundances obtained using $T_{\rm e}$(\siii) remains unexplored. So we believe the abundances from DM that have the determined temperature exceeding 5000~K from that expected using the  $T_{\rm{e}}$-$\rm{N_{2H\beta}}$ relation from \citet{2016MNRAS.457.3678P} are not reliable, and hence we discarded all these regions from the DM. 
We show the rejected regions as grey dots in  Fig. \ref{fig:direct_strong}. When taking into account only the black points, the dispersion of the DM around the S calibrator is $\sim0.1~\rm{dex}$, which is the expected value for SLM calibrations.

\begin{figure}
\includegraphics[width=1.0\linewidth]{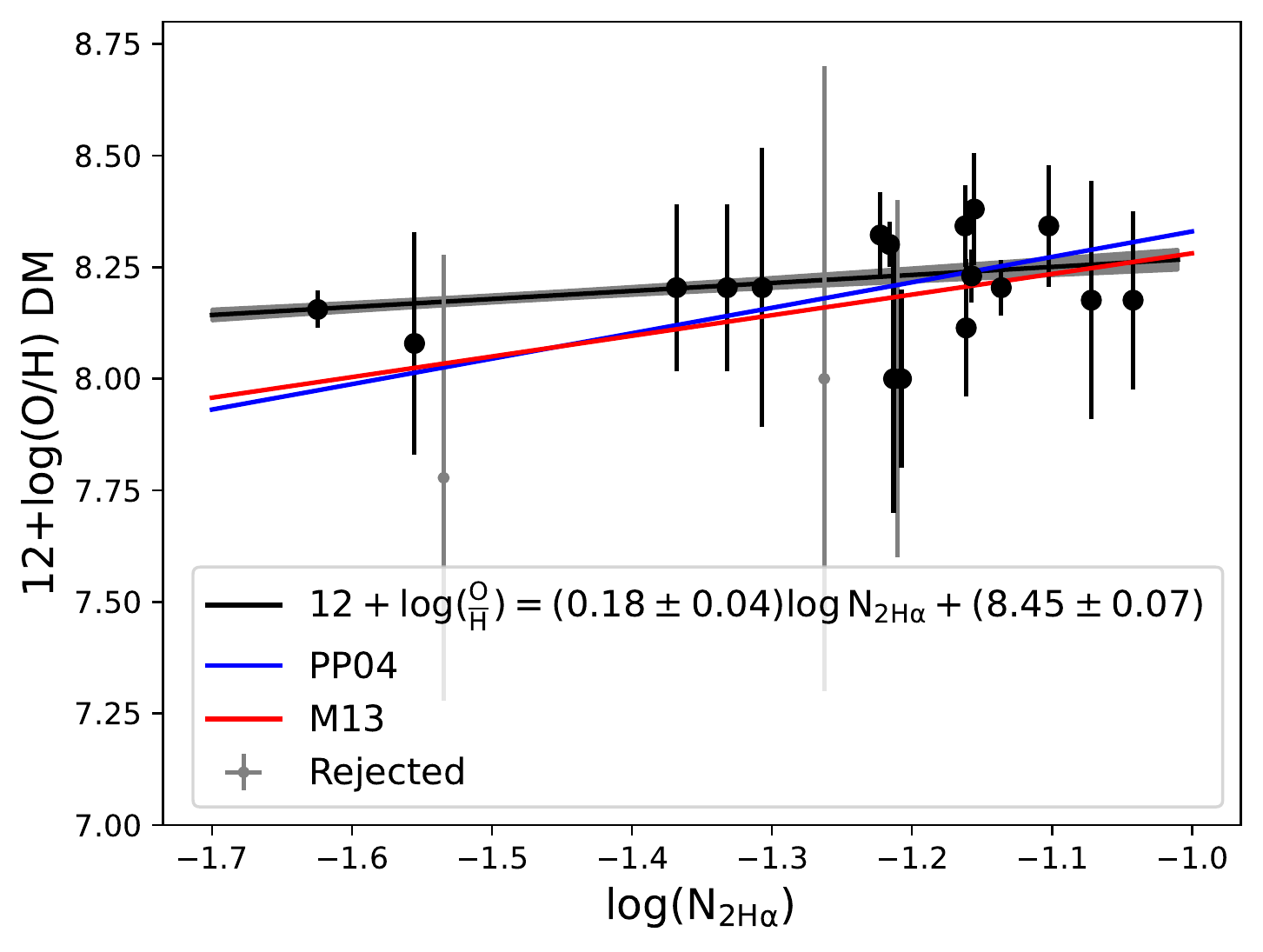}\\
\caption{O abundance 12+log(O/H) determined with the DM versus the 
$\rm{N_{2H\alpha}}=\rm{\frac{[NII]\lambda6548+\lambda6584}{H\alpha}}$ 
index for regions from the outer ring of Cartwheel. 
We plot as solid line our empirical calibration as the error weighted fit; grey points are rejected due to their large relative error. We include the calibrations from \citet{2004MNRAS.348L..59P} (PP04) and  \citet{2013A&A...559A.114M} (M13) as blue and red lines, respectively.}
\label{fig_n2cal}
\end{figure}

In addition, we have estimated our own calibration using the abundances from DM and the value of  $\rm{N_{2H\alpha}}=\rm{\frac{[NII]\lambda6548+\lambda6584}{H\alpha}}$. 
We show in Fig. \ref{fig_n2cal} the O abundance versus the $\rm{N_{2H\alpha}}$ ratio for the regions in the outer ring. We estimated  our O abundance empirical calibration performing a linear error weighted fit:
\begin{equation}
12+\log(\frac{\rm{O}}{\rm{H}})= (0.18\pm0.04)\log(\rm{N_{2H\alpha}})+(8.45\pm0.07) 
\label{eq_n2cart}
\end{equation}
The resulting fit is slightly flatter than the N2 calibrations of \citet{2004MNRAS.348L..59P} (N2-PP04), \citet{2013A&A...559A.114M} (N2-M13).

In order to compare between different empirical calibrations with the fiducial one (S  calibrator) and with the one obtained in this work for the regions in the outer ring of Cartwheel (N2-Cartwheel), we also estimate the O abundance using the calibrations from
N2-PP04 and N2-M13.
The O abundances determined using different empirical calibrations are discussed below.

\section{Discussion}

 We determined O, N, Fe and He abundances 
with the DM for 20, 20, 17, and 9 regions, respectively; with
 the temperature-sensitive line \siiia detected.
These regions are located in the outer star-forming ring.
  The median values for the different abundances are $12+\log\rm{\frac{O}{H}}$=8.19$\pm$0.15,
 $\log\rm{\frac{N}{O}}=-$1.57$\pm$0.09,  $\log\rm{\frac{Fe}{O}}=-$2.24$\pm$0.09, and  $12+\log\rm{\frac{He}{H}}$=10.90$\pm$0.03.
 The relative abundance of $\rm{\frac{N}{O}}$ is commonly approximated by $\rm{\frac{N}{O}}\approx\rm{\frac{N^{+}}{O^{+}}}$.
We corroborate that our estimate $\log\rm{\frac{N^+}{O^+}}=-$1.56$\pm$0.09 is consistent with the median value of $\rm{\frac{N}{O}}$ we found.

O, N and Fe are ejected into the ISM over different timescales, which allows us to use the measured abundances to determine the nature of the expanding ring wave of star formation in the Cartwheel. The O is mainly enriched in the ISM during the explosive death of massive stars during the core-collapse SNe, which also produce some amount of Fe \citep{1978MNRAS.185P..77E,1988BAAS...20.1100P,1990ApJ...363..142G}. Slightly prior to this event, the massive stars ($\geq25$~\msol) during their Wolf-Rayet (WR) phase ($\sim2-4$~Myr) can also enrich He, N and C due to the high mass loss ($\sim10^{-4}-10^{-5}$~\msolyr) through strong stellar winds \citep{Crowther2007, Maeder1992}. 
The N has two possible origins, either in massive stars in a ``secondary way'' using the C and O already present in the star at birth, or in a ``primary way'' using the freshly produced C and O and dredged up during the AGB phase \citep{Renzini1981, Villas1993}. On the other hand, most of the Fe in the ISM comes from the Type~Ia SNe over slightly longer timescales \citep{Matteucci1986}.

The WR star phase and core-collapse SNe explosions occur during the typical lifetime of \hii\ regions and hence nebular O, Fe, N and He abundances can have contribution from the recently expelled elements. This fact has motivated several authors to look for local enrichment of He/H and N/O ratios in \hii\ regions \citep[see e.g.,][]{Pagel1986,Pagel1992,Kobulnicky1997}. The results of these searches have been mixed so far with reports of local enrichment in the nearby starburst NGC\,5253 \citep{Lopez2007} and in a sample of WR galaxies \citep{Brinchmann2008} and inconclusive evidence in Mrk\,178 \citep{Kehrig2013}. 
\citet{Tenorio1996} argued that the expelled material in the stellar winds and SN ejecta could be trapped in hot coronal gas that would take $\sim10^{8}$~yr to cool down enough and allow an efficient mixing of the fresh elements with their local environment. The reason for non-enrichment could also be because some of the processed material can escape the star-forming zone during the SN explosions \citep{Recchi2004}.
Under these scenarios, the observed nebular abundances trace the metallicity of the ISM before the formation of the current generation of massive stars. In fact, the abundance could be even less than that of the pre-collisional disk if there is infall of metal-poor gas \citep{Koppen2005}.

\subsection{Fe/O, N/O vs O/H abundances}

\begin{figure*}
\begin{center}
\includegraphics[width=1.0\linewidth]{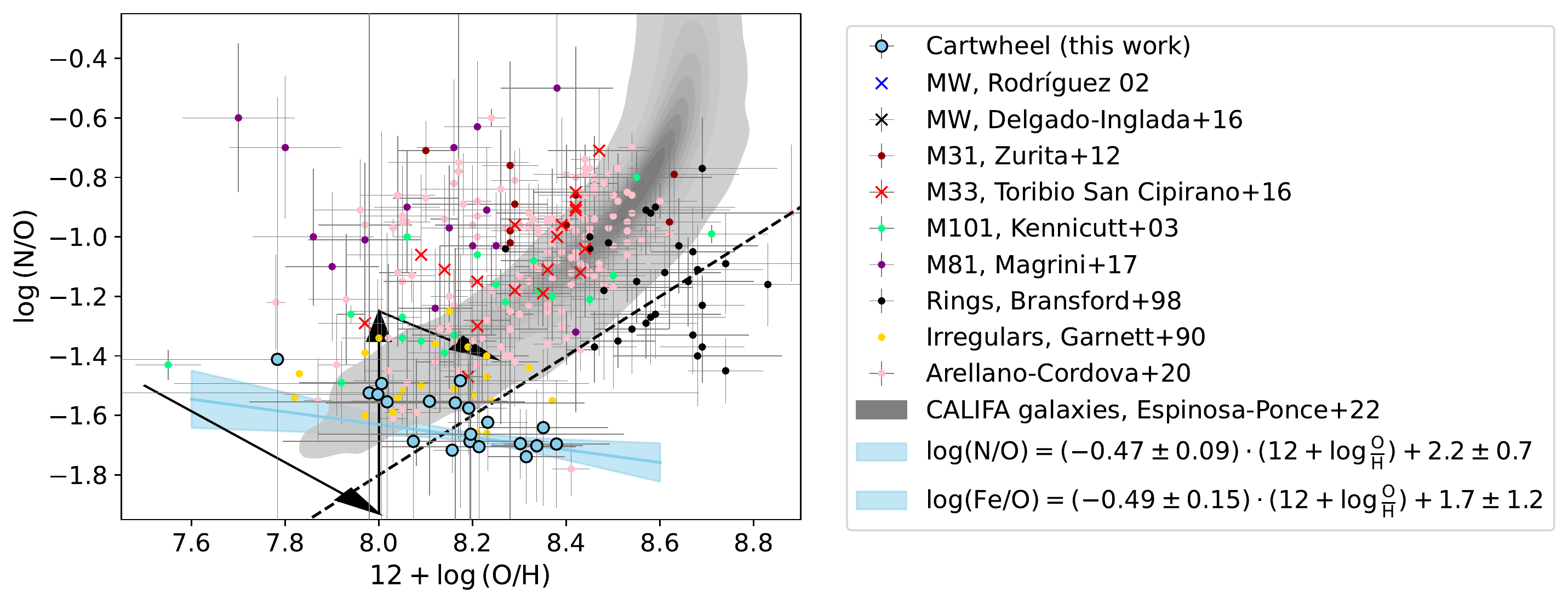}\\
\includegraphics[width=.5\linewidth]{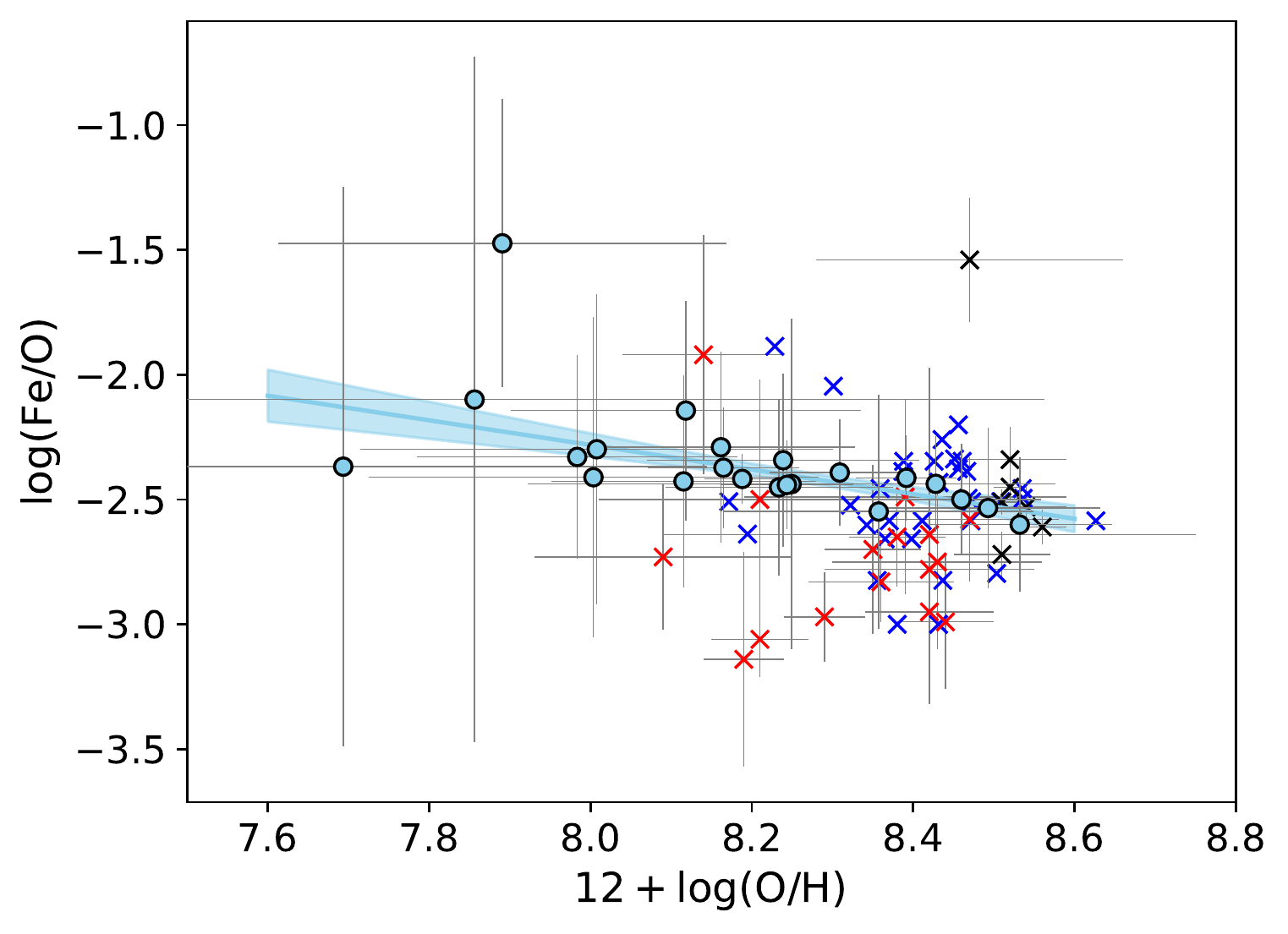}\\
\end{center}
\caption{log(N/O) (top) and log(Fe/O) (bottom) vs 12+log(O/H) derived in this work for the Cartwheel ring \hii\ regions using the DM (blue circles) compared with the corresponding values for \hii\  regions in nearby irregular and galaxies compiled from the literature: Irregulars \citep[yellow dots]{1990ApJ...363..142G};
M\,101 \citep[green dots]{2003ApJ...591..801K};
M\,31 \citep[dark red dots]{2012MNRAS.427.1463Z};
M\,33 \citep[red crosses]{2016MNRAS.458.1866T};
M\,81 \citep[purple dots]{2017MNRAS.464..739M}; 
regions of collisional ring galaxies \citep[black dots]{Bransford1998}
and regions from different nearby galaxies in \citep{Karla2020} (pink dots) and 
 \citep{2020MNRAS.494.1622E} (grey area). 
The vectors and the dashed line shown in the top panel are taken from the schematic representation of the evolution of the N/O ratio for a dwarf galaxy from \citet{1990ApJ...363..142G};
 The bottom panel includes measurements for the Milky Way (MW)  
 \hii\ regions \citep[blue and black crosses, respectively]{2002A&A...389..556R,2016MNRAS.456.3855D}.  
 The blue lines are the error weighted linear fits to the 
 Cartwheel data points, while the shaded area represents 
 the $1-\sigma$ 
 uncertainty range of the fits.
 }
\label{fig:ab_direct}
\end{figure*}

Oxygen produced in stars is returned to the ISM and mixed with the ambient cooler gas 
over relatively short time scales (a few$\times10^7$~yr) as compared to the enrichment of N and Fe (a few$\times10^8$~yr). Rings of ring galaxies fade over a time scale of a few$\times10^8$~yr. In the classical ``expanding density wave'' models of ring galaxies, the location of star-forming ring moves outward in time, leaving behind the products of SF, i.e. the recently processed elements, at inner radii \citep[e.g.][]{Korchagin2001}. The presently observed star forming ring contains stars formed over the past 10~Myr only, which is too short a period for the enrichment of even O. Hence, the ring \hii\ regions are expected to trace the metallicity of the pre-collisional disk at the current location of the ring \citep[see e.g.][]{Bransford1998}. The metallicity at the ring could be even less than that of the pre-collisional disk if there was an infall of metal-poor gas during the collisional event that triggered the ring formation. 

 \citet{Renaud2018}, who took into account gas dynamics at parsec scales in their recent simulation of ring galaxies, find results that differ qualitatively and quantitatively from the expectations of the classical ``expanding density wave'' scenario.
The gas is swept by the rapidly expanding wave which takes along with it all the products of SF.  Recent simulations by \citet{2021NGC922} also support this idea. Under this ``expanding material-wave'' scenario, the ring contains most of the elements formed in the expanding wave. Furthermore, this model predicts timescale of $\sim$100~Myr for the ring in the Cartwheel. This timescale is long enough for the expulsion and enrichment of O in the ISM from the stars formed in the wave, but short enough for the enrichment of primary N and Fe produced in SNIa \citep{Recchi2006}. The production of the secondary N in massive stars becomes efficient at $12+\log\frac{\rm{O}}{\rm{H}}\ge8$ \citep{Vincenzo2016}, which suggests that the highest metallicity Cartwheel \hii\ regions could have a relatively higher N/O ratio.

The trends in metallicity expected under the two scenarios of the ring formation described above are different. In the ``expanding density wave'' scenario, enrichment of O due to SF triggered by the ring wave is expected to occur throughout the disk, with some enrichment of N and Fe at the inner most radii. The ring is not expected to show any metal inhomogeneity. Under this scenario, the variation in the O abundance in the ring \hii\ regions can only arise due to different amount of metal-poor infalling gas in different parts of the ring. However, gas infall would not change the N/O and Fe/O ratios and hence these ratios are expected to be independent of the O/H variation. On the other hand, in the ``expanding material-wave'' scenario, each \hii\ region contains gas that was enriched by metals expelled from stars over all radii, and hence ages, since the wave started to expand. 
An azimuthal zone that received material from intense past bursts is expected to have higher O/H as compared to a zone which has not received any metal-enriched gas. The N/H and Fe/H values in the ring \hii\ regions are not expected to deviate much from their pre-collisional values because of the time ($\sim$100~Myr) elapsed since the ring-making collision is shorter than the typical time scale required to enrich N and Fe  through the primary process. Thus, N/O and Fe/O are expected to systematically decrease as O/H increases. The slope of the relation would depend on the relative contribution of the  secondary process in massive stars to the production of N and Fe, with the slope close to unity if the entire N and Fe  is of secondary origin in stars less massive than 8~\msol.  The production of the secondary N in massive stars becomes efficient at $12+\log\frac{\rm{O}}{\rm{H}}\ge8$ \citep{Vincenzo2016}, which suggests that the highest metallicity Cartwheel \hii\ regions could have a relatively higher N/O ratio.

We show the N/O (top) and Fe/O ratios (bottom) versus 12+log(O/H) abundances for all the \hii\ regions with $T_{\rm e}$ measurements in Fig.~\ref{fig:ab_direct}. We see a systematic decrease of both N/O and Fe/O with O/H, which is inconsistent with the trend expected from the ``expanding density wave'' scenario, but in agreement with the trend expected from the alternative scenario of the ``expanding material-wave''. The slopes of the linear fits to the observed relation (blue lines) are close to 0.5, which suggests non-negligible  secondary contribution to N and Fe production from stars more massive than 8\msol\ in the Cartwheel.

In Fig.~\ref{fig:ab_direct}, we also compare the location of the Cartwheel ring \hii\ regions with that of \hii\ regions in other galaxies. The comparison sample for N/O includes \hii\ regions in nearby spiral galaxies, irregular galaxies and \hii\ regions in collisional ring galaxies analysed by \citet{Bransford1998}. The Cartwheel \hii\ regions span  $\sim$0.6~dex range in O/H at the low-metallicity end. The N/O values in the Cartwheel \hii\ regions are among the lowest values in the comparison sample.
The trend of N/O ratio in the Cartwheel  is consistent  with the trend seen for irregular galaxies and a few regions in M101. The rest of the galaxies, including the relatively metal-rich ring galaxies studied by \citet{Bransford1998}, show N/O enhancements by as much as 0.5--1.0~dex over the values found for the high-metallicity Cartwheel regions. This suggests that the outer disk of the 
pre-collisional Cartwheel is chemically less evolved as compared to other ring galaxies.

Data on the Fe abundance for extra-galactic \hii\ regions are scarce beyond the Local group, given the faintness of the Fe lines in \hii\ regions. Hence, we compare the results with \hii\ regions in M\,33 \citep{2016MNRAS.458.1866T} and the Galactic \hii\ regions \citep{2002A&A...389..556R, 2016MNRAS.456.3855D}. These data cover Fe/O ratios only for 12+log(O/H)$\gtrsim8.05$. The Fe/O ratio reported here for the Cartwheel galaxy is in agreement with those observed in the Galactic \hii\ regions and M\,33.
Unfortunately, there are no measurements of Fe/O in low metallicity \hii\ regions in nearby galaxies to compare with.

\subsection{Azimuthal variation of O using SLM}

\begin{figure*}
\begin{centering}
\includegraphics[width=0.495\linewidth]{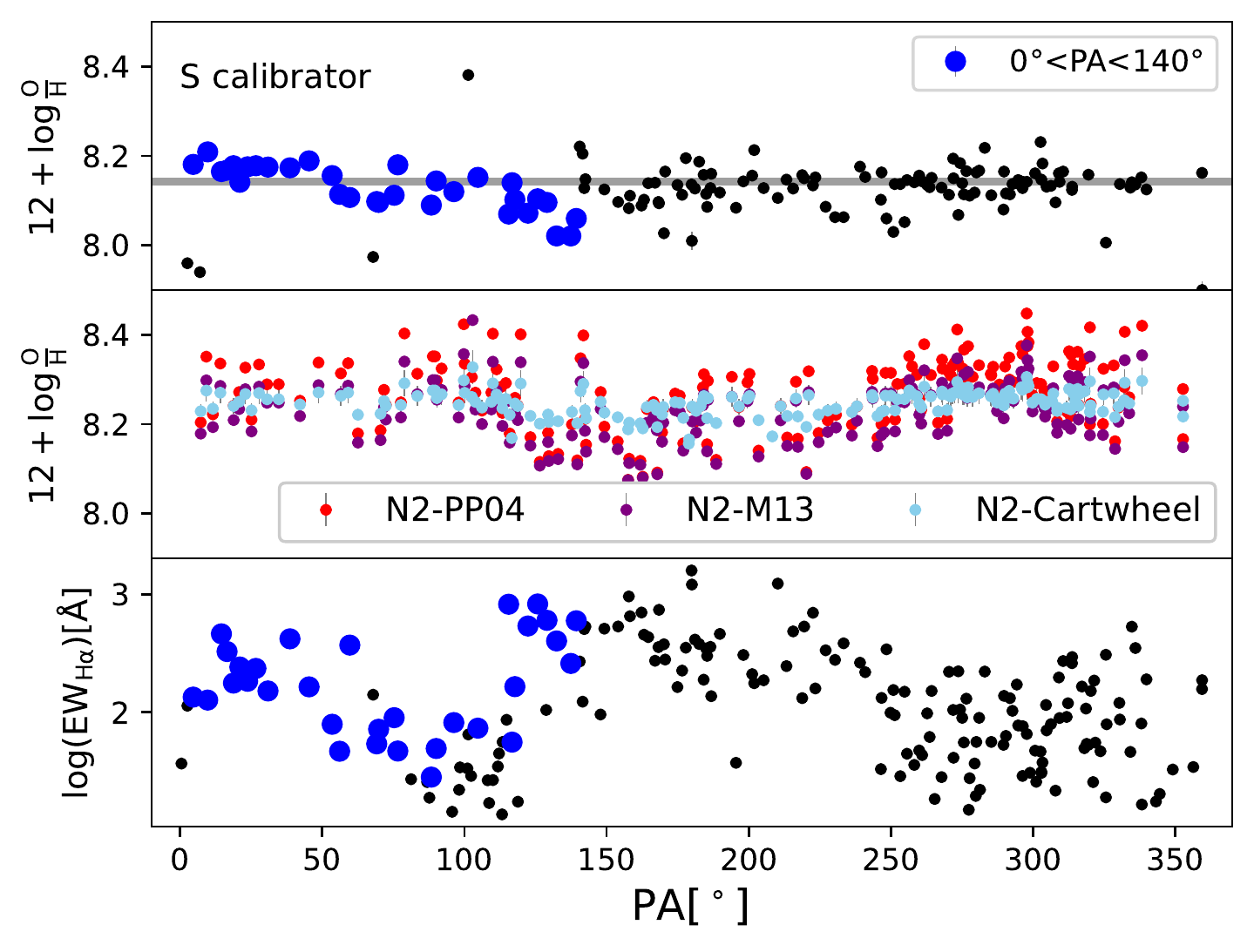}~
\includegraphics[width=0.495\linewidth]{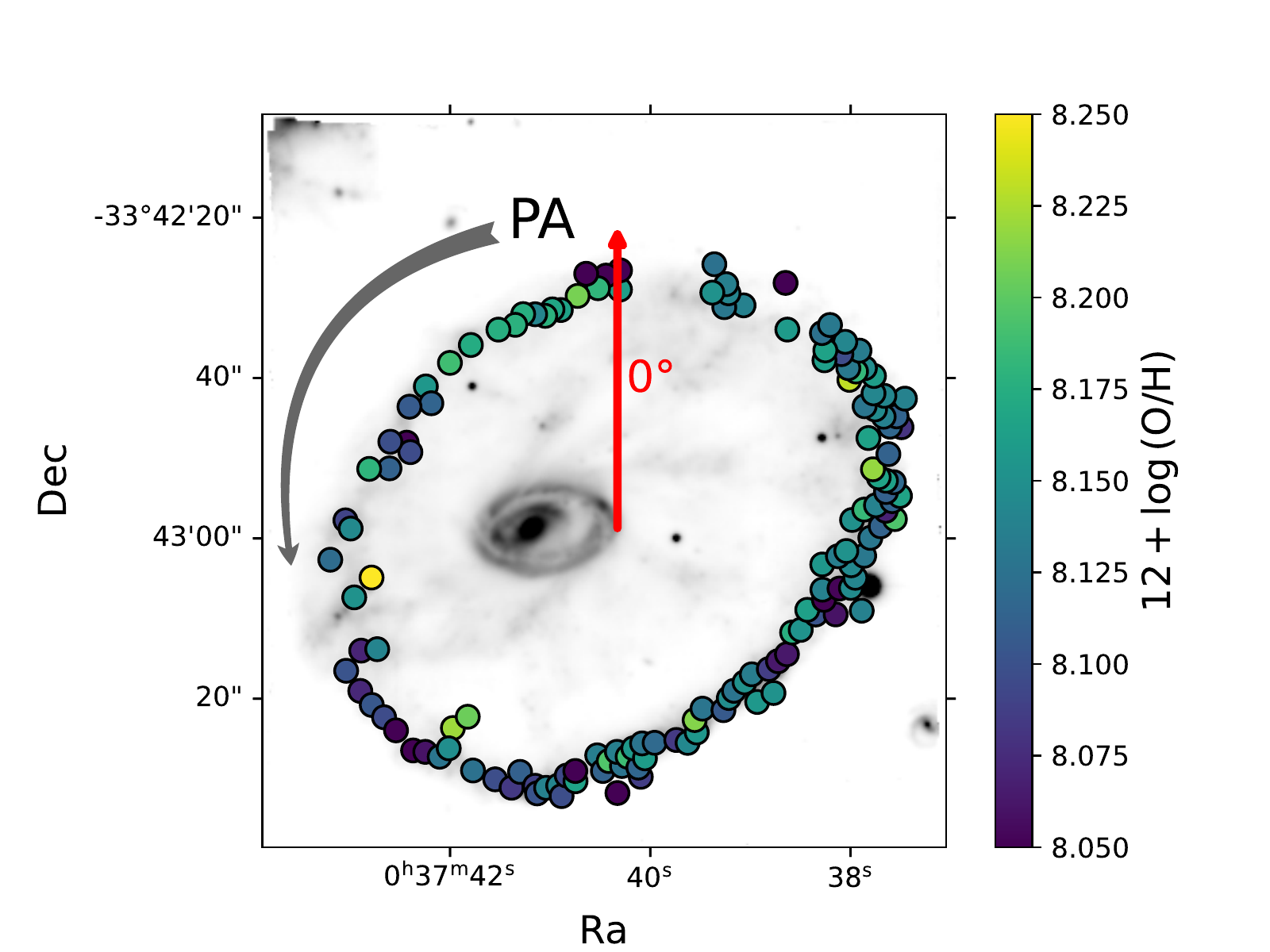}\\
\par\end{centering}
\caption{ (left) O abundances using the S-calibrator vs PA in the external ring of Cartwheel. The azimuthal behaviour for other SLMs are shown in the middle left panel:  N2-PP04 (red), N2-M13 (purple) and N2-Cartwheel (light blue). We also plot the  EW(H$\alpha$) versus PA (bottom panel). Regions between 0 and 140 degrees in PA are shown in blue.
The grey horizontal line is the median O abundance value for the S calibrator.
(right) SDSS r-band image of the Cartwheel indicating the O abundance values derived with the S calibrator  (colour points). The grey arrow indicates the positive direction starting from PA=$0^{\circ}$ (red arrow). 
}
\label{fig:abun_pa}
\end{figure*}

\begin{figure}
\includegraphics[width=1.\linewidth]{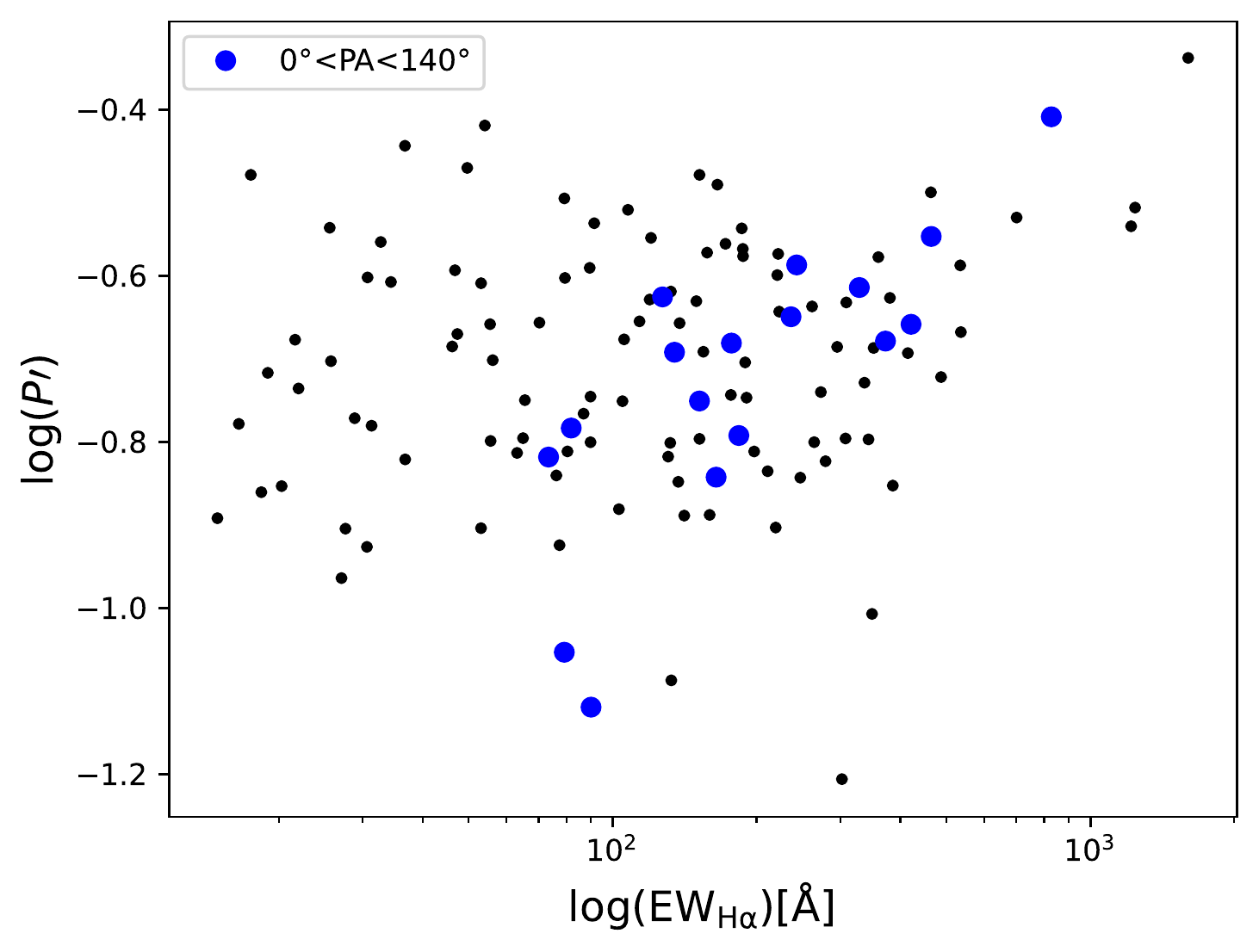}\\
\caption{
Excitation parameter ($P^{\prime}$) vs EW(H$\alpha$). Regions between 0 and 140 degrees in PA (as seen in Fig. \ref{fig:abun_pa}) are displayed in blue. 
}
\label{fig:ppparam}
\end{figure}

The star-forming ring of the ring galaxies shows considerable azimuthal variation in the distribution of intensity in most bands \citep[see e.g.,][Fig.5]{Madore2009}. The Cartwheel is no exception, with the southern half of the ring being noticeably brighter in all bands \citep[see e.g.][]{Marcum1992, Mayya2005}. This brightness is due to enhanced SF in this part of the ring \citep{Higdon1995}. 
The distribution of the atomic and molecular gas also shows considerable azimuthal structures \citep{Higdon1996, Higdon2015}. The azimuthal asymmetry is believed to be due to the non-central impact point of the intruder \citep[e.g.][]{Renaud2018}.

It is of interest to explore the azimuthal variation of O abundances of the ring \hii\ regions. With this purpose, we plot O abundance as a function of PA of the \hii\ regions in the ring in Fig.~\ref{fig:abun_pa} (left). For measuring the PA, we used the center of the ellipse that best matches the star-forming ring (ellipse \#9 in Fig.~\ref{fig:ellipses}) instead of the nucleus.
This center, $\alpha=0\rm{h}37\rm{m}40.27\rm{s}$; $\delta=-33\rm{^{o}} 42\rm{'}59.06\rm{''}$, is illustrated on the SDSS r band image extracted from the MUSE cube in the right panel of Fig. \ref{fig:abun_pa}. The O abundance values using the S calibrator are shown in colour for all the ring \hii\ regions using the code in the colour-bar on the right.

The O abundance for the majority of the \hii\ regions using the S-calibrator lies close to the median value of  12+log(O/H)=8.14 (grey line in the top-left panel of Fig.~\ref{fig:abun_pa}), with a spread of $\sim$0.1~dex, which is the typical error of the calibration. However, in the PAs between 0$^\circ$ and 140$^\circ$, a systematic azimuthal gradient is noticeable. 
 The $12+\log\frac{\rm{O}}{\rm{H}}$ value decreases gradually by 0.2~dex in this  140$^\circ$ azimuthal range. We indicate these regions as blue circles in Fig.~\ref{fig:abun_pa} (top-left). In order to explore whether the observed azimuthal gradient is real, or an artefact of the use of SLM, we plot in the middle panel the O abundance calculated using other three SLM
calibrators: N2-Cartwheel (light blue; this work), N2-PP04 (red), and N2-M13 (purple). None of these calibrators show the systematic variation between 0$^\circ$ and 140$^\circ$. Thus, it is likely that the systematic variation using the S-calibrator is due to some other physical effect that affects the S-calibrator more than other calibrators. The \ha\ EW plotted in the bottom-left panel also shows a systematic decrease in the same azimuthal range. The \ha\ EW decreases with age \citep{1999ApJS..123....3L} and hence the high EW regions are expected to be systematically younger by $\sim$5--10~Myr as compared to low EW regions. There is no reason to believe these relatively older regions are metal-poor as compared to their younger counterparts. On the other hand, the ionization parameter decreases with age, and hence the observed $\sim0.2$~dex decrease of the O abundance in the 0--140$^\circ$ azimuthal range could be due to the inadequacy of the S-calibrator over the entire observed range of the ionization parameter. We indeed observe a systematic decrease of the ionization parameter, defined as\footnote{We labeled the ionization parameter as $P^\prime$ to distinguish from the more commnonly used
$P=\frac{I(\rm{O[III]\lambda4959+5007})}{I(\rm{O[III]\lambda4959+5007})+I(\rm{O[II]\lambda3727+29})}$, which cannot be defined for the MUSE dataset due to the non-coverage of the $O[II]\lambda3727+29$ line.}, $P^{\prime}=\frac{I(\rm{S[III]\lambda9069})}{I(\rm{S[III]\lambda9069})+I(\rm{S[II]\lambda6717+31})}$, in the azimuthal range (PAs=0--140$^\circ$) where the \ha\ EW decreases. 
In Fig.~\ref{fig:ppparam}, we plot $P^{\prime}$ vs \ha\ EW for all the ring regions. The  $P^{\prime}$ for regions in the 0--140$^\circ$ azimuthal range (shown as blue dots) clearly correlates with \ha\ EW. This confirms that the S-calibrator we have used does not take into account the age-dependent variation of the ionization parameters in the Cartwheel \hii\ regions. It seems the N2-calibrators are less affected by the cluster evolution. We ignore any intrinsic correlation between ionization parameter and metallicity in \hii\ regions \citep[e.g.][and references therein]{Kewley2019}, and conclude that we do not have compelling evidence for azimuthal variation of O abundance in the Cartwheel.

\subsection{Radial abundance gradient of O}

\begin{figure}
\begin{centering}
\includegraphics[width=1.0\linewidth]{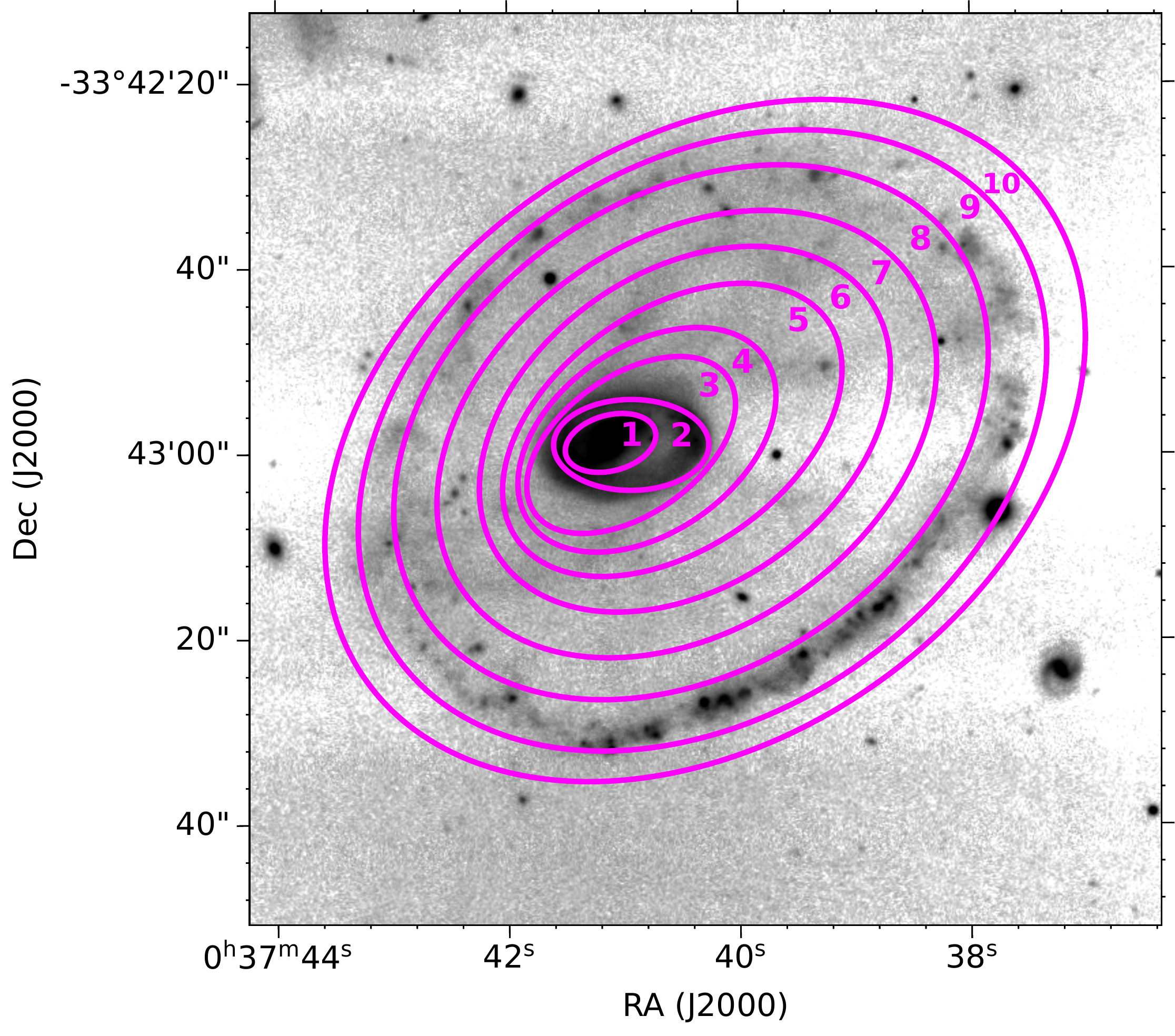}
\par\end{centering}
\caption{K band image from \citet{Barway2020} of the Cartwheel galaxy. We show the ten different ellipses (magenta) used to estimate the radius depending on the position in the Cartwheel galaxy.
}
\label{fig:ellipses}
\end{figure}

    SLM empirical calibrations  allowed us to trace the O abundances of the region between the nucleus and the outer ring, which enables for the first time, the determination of the abundance gradient in the disk of the Cartwheel. Obtaining the abundance gradient requires the estimation of galactocentric distances for each \ha-emitting region. Under the collisional scenario, ring galaxies have two centers: one defined by the position of the nucleus;  the other by the impact point of the intruder galaxy, defined as the center of the ellipse that best fits the star-forming ring. Pre- and post-collisional gradients, if present, are expected to be symmetric around the nucleus and the impact point, respectively. In order to measure the colour gradients in the Cartwheel, \citet{Marcum1992} proposed a scheme wherein they changed the center of symmetry smoothly from the nucleus at small radii to the impact point at large radii. We followed their scheme and defined ten different ellipses, which are shown in Fig.~\ref{fig:ellipses}. The PA and ellipticity of the two innermost ellipses are dictated by the bar and the inner ring. For the subsequent ellipses, the PA is fixed at the value corresponding to that of the outer ring, whereas the ellipticity was decreased gradually from the inner to outer ellipses. We took the semi-major length of the ellipse as the galactocentric distance of the corresponding ellipse. For each region where we measured the O abundance using the SLM, we determined its galactocentric distance by assigning the parameters of the ellipse that passes through the region.

\begin{figure}
\begin{centering}
\includegraphics[width=1.0\linewidth]{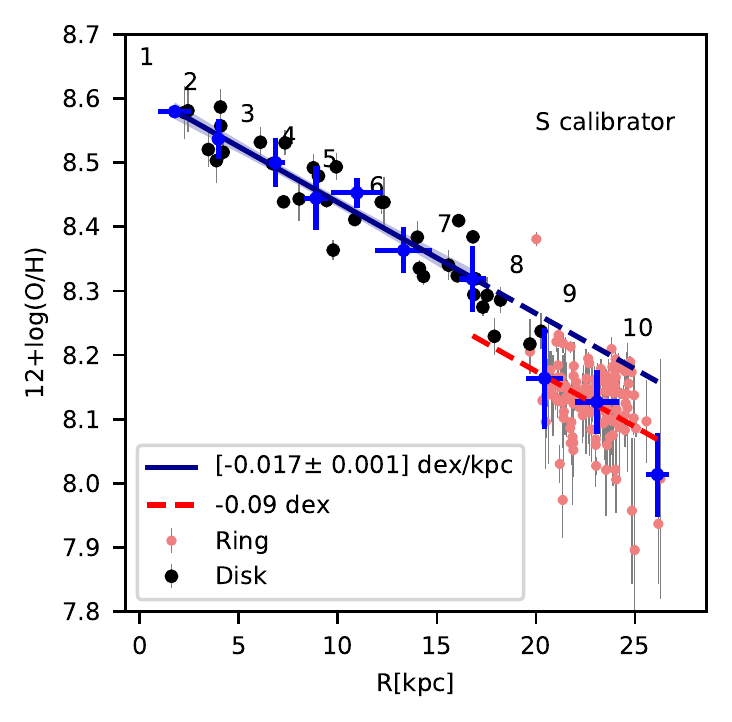}

\par\end{centering}
\caption{ $12+\log(\rm{O/H})$ vs the radius (R) for the fiducial empirical calibration: S-calibrator \citep{2016MNRAS.457.3678P}.
Black points are regions identified in the disk, while red points are those identified in the ring. The blue points are the average values for the regions inside each ellipse (labelled with the number of the ellipse), while the blue line is the fit to the average values of the first seven ellipses and the blue shaded region is the 1-$\sigma$ uncertainty range of the fit. We show as a blue dashed line the extrapolation of the linear fit to larger radius, while as a red dashed line the same line, but shifted downwards by $-0.09$~dex.
}
\label{fig:oxgrad1}
\end{figure}

We plot the radial distribution of the O abundances in Fig. \ref{fig:oxgrad1}, where we show each region in the disk as black points and those in the ring as red points.  We plot the radial distribution for the fiducial empirical calibration mentioned above:
S-calibrator. 
We also show the average values for the regions between two consecutive ellipses by blue crosses, where the horizontal and vertical sizes represent the standard deviation of points enclosed by each of our ten ellipses.  Clearly-defined abundance gradient can be seen in the zone between the inner and the outer ring. We have fitted the average values (blue crosses) of the first seven ellipses by a straight line, which is shown as a continuous blue line. The measured abundances of almost all the regions in the outer ring are systematically lower than the value expected from the extrapolation of the fit, which is shown by a blue dashed line. We find that the abundances of ring regions are consistent with the slope of the fit to the disk regions, but shifted downwards by $-0.09$~dex.  
The value of the gradient is   $-0.017\pm0.001\thinspace\rm{dex/kpc}$. 
Assuming  $R_{25}=24.95$~kpc \citep[{HyperLeda database, }][]{2014A&A...570A..13M}, which is just at the outer edge of the outer ring, the gradient is equivalent to a value of   $-0.42\pm0.03$\thinspace{dex/$R_{25}$}.

\begin{figure}
\begin{centering}
\includegraphics[width=1.0\linewidth]{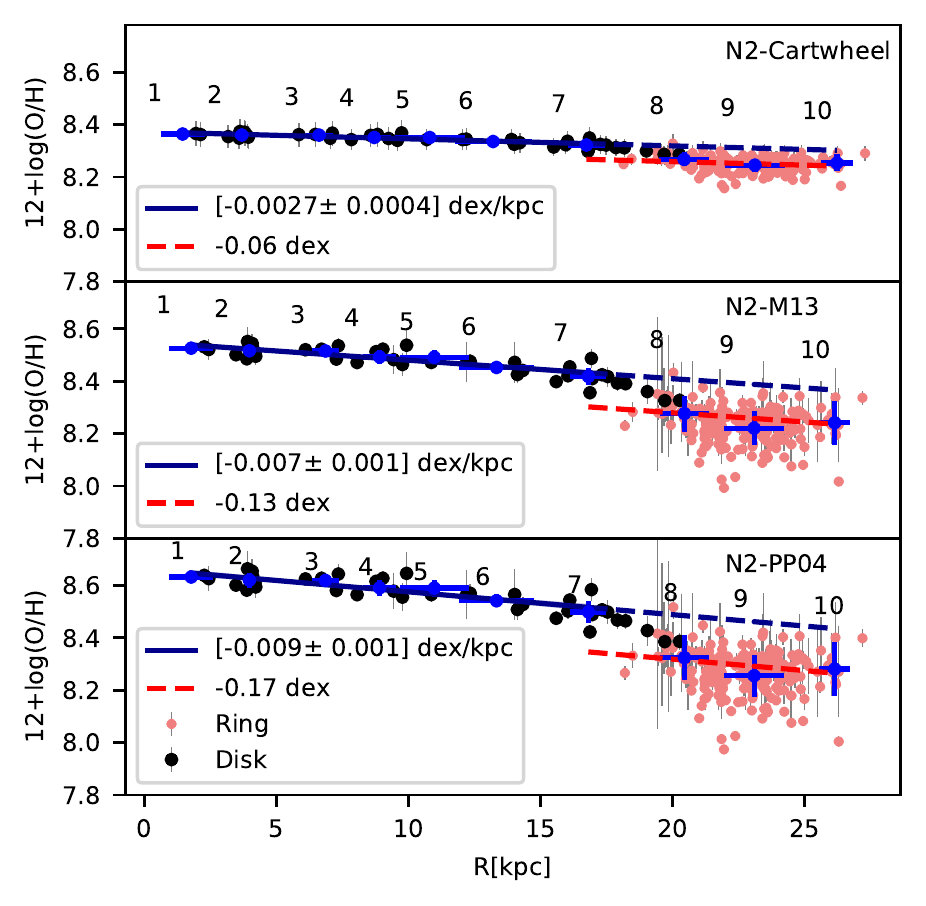}

\par\end{centering}
\caption{$12+\log(\rm{O/H})$ vs radius (R).  We plot the O abundance for three empirical calibrations:  N2-Cartwheel (top panel; this work); N2-M13 \citep[middle; ][]{2013A&A...559A.114M} and N2-PP04 \citep[bottom; ][]{2004MNRAS.348L..59P}.
Black points are regions identified in the disk, while red points are those identified in the ring. The blue points are the average values for the regions inside each ellipse (labelled with the number of the ellipse), while the blue line is the fit to the average values of the first seven ellipses and the blue shaded region is the 1-$\sigma$ uncertainty range of the fit. We show as a blue dashed line the projection of the linear fit to larger radius, while as a red dashed line the same projection although shifted downwards by  factor which depends on the empirical calibration used. 
}
\label{fig:oxgrad2}
\end{figure}

 Since O abundance can depend on the assumed empirical calibration, we also show in Fig. \ref{fig:oxgrad2} the gradient for the three different empirical calibrations discussed above: N2-Cartwheel (upper; this work); N2-PP04 (middle), and N2-M13 (bottom).
 The two basic characteristics found using the S-calibrator, namely a negative abundance gradient and the downward shift in abundance in the ring, are seen for all the three N2-calibrator relations.
The values of the gradient and the shift obtained for the N2-M13 and N2-PP04 calibrators agree within each other, with mean values of gradient=$\sim-0.008$~dex/kpc and shift=$-0.15$~dex. This gradient is around a factor of two shallower and the shift 66 percent larger, as compared to that of the S-calibrator. 
The N2-Cartwheel calibration has been obtained using the regions in the outer ring. The extrapolation of this calibration to the regions in the inner disk is probably  incorrect due to the systematically lower ionization parameter of the inner disk regions, which is most likely the reason for its discrepancy with respect to the other two N2-calibrators.

The difference between methods is consistent with that reported by \citet{2021MNRAS.500.2359Z} for 2831 HII regions in nearby galaxies.

The slope using the fiducial calibrator is steeper compared with those of normal galaxies reported in \citet{2019A&A...623A.122P} ($-0.158$\thinspace dex/$R_{25}$). Assuming an effective radius of $r_{\rm{e}}=5.13\rm{kpc}$  \citep[J band][]{2006AJ....131.1163S},
the slope is  $-0.087\pm0.005\thinspace\rm{dex/r_e}$ for the S calibrator.
The gradient is in agreement with those found for normal galaxies from different surveys such as CALIFA  \citep[$-0.07\pm0.05\thinspace\rm{dex/r_e}$][]{2016A&A...587A..70S}, MANGA \citep[$-0.07\pm0.1\thinspace\rm{dex/r_e}$][]{2017MNRAS.469..151B}, and AMUSING \citep[$-0.1\pm0.03\thinspace\rm{dex/r_e}$][]{2018A&A...609A.119S}, see \citet{Sanchez2020} for a review. 
The gradients estimated using the N2 calibrators are slightly shallower, as those normally  found in interacting galaxies 
 \citep{2010ApJ...723.1255R,2014A&A...563A..49S}.
 However, since we are dealing with regions with very different ages and excitation parameters in the inner disk and the outer ring, the gradient value obtained using the fiducial S-calibrator is more trustworthy.

The observed agreement of the gradient in the Cartwheel with that in 
the normal galaxies over the characteristic radii 
$R_{25}$ and $r_{\rm{e}}$ would suggest that the star 
formation in the disk of the Cartwheel following the 
ring-producing interaction has not resulted in noticeable 
change in its abundance properties in the inner disk, 
assuming the  $R_{25}$ and $r_{\rm{e}}$ have not 
undergone changes due to interaction.
 The downward shift of the O abundance in the ring regions is present (although with different values) in all the calibrations. Therefore, the existence  of an abundance shift in the outer ring is rather convincing.

\subsection{Stellar mass surface density vs metallicity relation}

Metals expected from stars formed by the passage of the expanding density wave are expected to contribute to the creation of metallicity gradient \citep[e.g.][]{Korchagin1999}.
However, to establish whether or not the observed gradient in the Cartwheel is created by the expanding wave, a knowledge of the gradient in its pre-collisional disk is required.
Metallicity in a galaxy is related to the mass in old stars, e.g.  \citet{Maiolino2019} and references therein.  
The stellar mass  for the Cartwheel was determined using the K-band total flux and a mass-to-light ratio appropriate to the disk of the Cartwheel. We used the K-band image from \citet{Barway2020} to obtain a total flux from old stars by masking all bright \hii\ regions. We obtained K=11.50, which is in good agreement with the measurement of K=11.55, reported by \citet{Marcum1992}. We used $\log(\frac{M}{L_K})=-0.2$ from \citet{Bell2003} corresponding to a disk colour of $B-V$=0.4 and ${M_K}_\odot$=3.33, to obtain a log(M$_*$/M$_\odot$)$=10.75$ for the Cartwheel. 
For this mass,
normal disk galaxies are expected to have a gradient of $\sim-0.01\thinspace\rm{dex/kpc}$ \citep{2019A&A...623A.122P}, which is marginally shallower than the gradient obtained for the Cartwheel. In order to explore further the origin of the gradient, we carry out a spatially resolved analysis of the stellar mass surface density ($\Sigma_{*}$) vs metallicity relation. A local  relation between these two quantities has been found in normal spiral galaxies \citep{2012ApJ...756L..31R}. 
 Anomalies from this local relation are observed and associated to gas accretion, for example due to to galaxy interactions \citep{2019ApJ...872..144H,2019ApJ...882....9S}.
We compare the average abundance for all the regions in the zone enclosed by successive ellipses with the azimuthally averaged $\Sigma_{*}$, determined from the corresponding K-band surface brightness and using the $\log(M/L_{\rm{K}})$ mentioned above.

\begin{figure}
\begin{centering}
\includegraphics[width=1.0\linewidth]{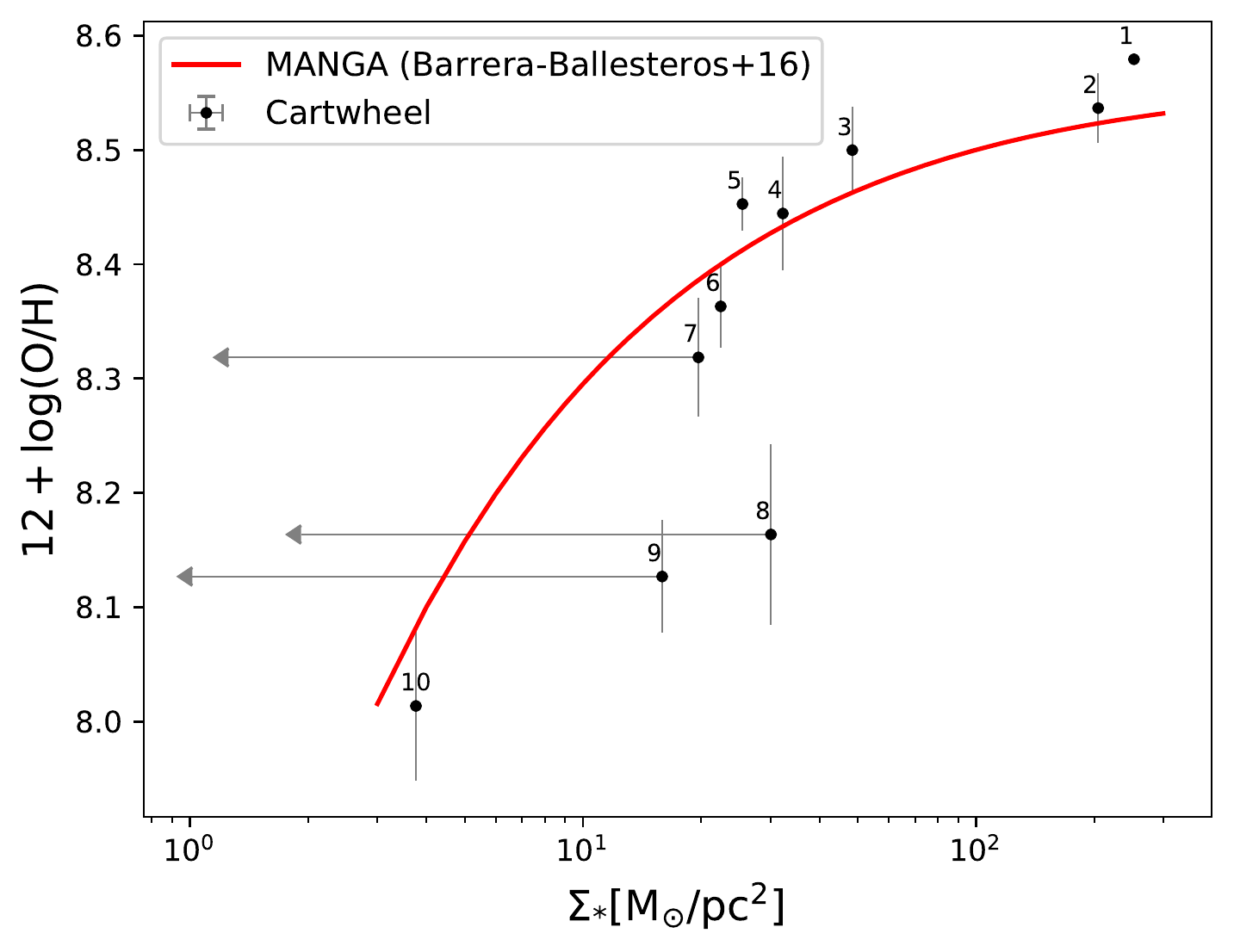}
\par\end{centering}
\caption{$12+\log(\rm{O/H})$ vs the stellar mass surface density ($\Sigma_{*}$). We show as black dots the points for the 10 ellipses defined in Cartwheel (labelled with the number of the ellipse). Ellipses 7, 8 and 9 are affected by young stellar component contamination. The arrows represent the maximum effect of this contamination. 
We show as a red line the metallicity-$\Sigma_{*}$ relation reported by the MANGA survey  \citep{2016MNRAS.463.2513B}.
}
\label{fig:ste_met}
\end{figure}

We compare $\Sigma_{*}$ with $12+\log(\rm{O/H})$ in Fig. \ref{fig:ste_met} for each of the elliptical annulus. 
We show the results for the Cartwheel galaxy as black stars and compare these points with the relation reported by the MANGA survey \citep[][]{2016MNRAS.463.2513B}. 
The points for the inner six ellipses and to some extent the outer-most ellipse, are in excellent agreement with $\Sigma_{*}$-metallicity relation for normal galaxies observed by the MANGA survey. On the other hand, the points for ellipses \#7, \#8 and \#9  are too metal-poor for their stellar content, or too bright for their observed (low) metallicity. Ellipse \#9  encloses the star-forming ring, where our assumption that all the K-band flux comes from old stars clearly overestimates the stellar mass. Ellipses \#8 and \#7 are in the wake of the expanding wave and hence the regions in these annuli are expected to contain descendants of massive stars, especially the red supergiants (RSGs) which could contribute substantially to the observed K-band surface brightness. We hence obtained an alternative value of stellar surface density assuming that the entire observed K-band surface brightness originates in RSGs. For this purpose we used a $\log(M/L_{\rm{K}})=-1.2$ from \citet{Eldridge2017} models for a 6~Myr population. 
These are indicated by the arrows. The real $\Sigma_{*}$ is expected to be somewhere between these two extreme cases, depending on the relative contribution of wave-induced star formation and pre-collisional disk stars to the K-band surface brightness. The sizes of the arrows are consistent with this scenario of contamination of the K-band SB of ellipses \#7, \#8 and \#9 from wave-induced SF and hence the observed slope of the metallicity gradient in the inner-most six ellipses and the zone outside the star-forming ring represents that of the pre-collisional disk.

\subsection{Chemical evolution scenario in the Cartwheel and other ring galaxies}

Combining the results of abundance gradient (Fig.~\ref{fig:oxgrad1}) with the  local  mass-metallicity relation (Fig.~\ref{fig:ste_met}) suggests that the observed abundance gradient in the Cartwheel is inherited from the pre-collisional disk rather than from metals expelled from stars formed by the passage of the expanding wave in the disk. 
On the other hand, the decrease of N/O and Fe/O with the O abundance implies the presence of recently processed O, but not much of N and Fe, in the ring \hii\ regions. This can be explained as due to the displacement of the O processed in massive stars to the current location of the ring, before the enrichment of  primary N and Fe. The age of $\sim$100~Myr inferred for the ring of the Cartwheel in recent simulations incorporating parsec-scale hydrodynamical processes by \citet{Renaud2018} is consistent with the chemically constrained age. 
 However, Fig.~\ref{fig:oxgrad1} suggests a clear downward offset in O abundance by $\sim$0.1~dex for the ring \hii\ regions with respect to the extrapolation of the gradient of the inner disk. If the ring contains O processed after the ring-making interaction, the O abundance is expected to be higher, rather than lower, than the pre-collisional value. This apparent contradiction can be understood as due to the mixing of the ISM containing the pre and post-collisional elements in the ring by the gas that is more metal-poor than that of the pre-collisional disk \citep[e.g.][]{Koppen2005}. One way of acquiring this metal-poor gas is through gas infall from the intruder, which has to be metal-poor. 
In the ``expanding material-wave'' scenario proposed by \citet{Renaud2018}, the processed elements are mixed naturally with the accreted metal-poor gas at successively outer radius. 
Most of the gas mass in the Cartwheel is in the atomic form \citep{Higdon1996, Higdon2015}, which is concentrated in the outer ring, which supports our hypothesis that dilution of ISM from metal-poor gas is responsible for the lower than the expected metallicity in the ring regions.

\begin{figure}
\begin{centering}
\includegraphics[width=1.0\linewidth]{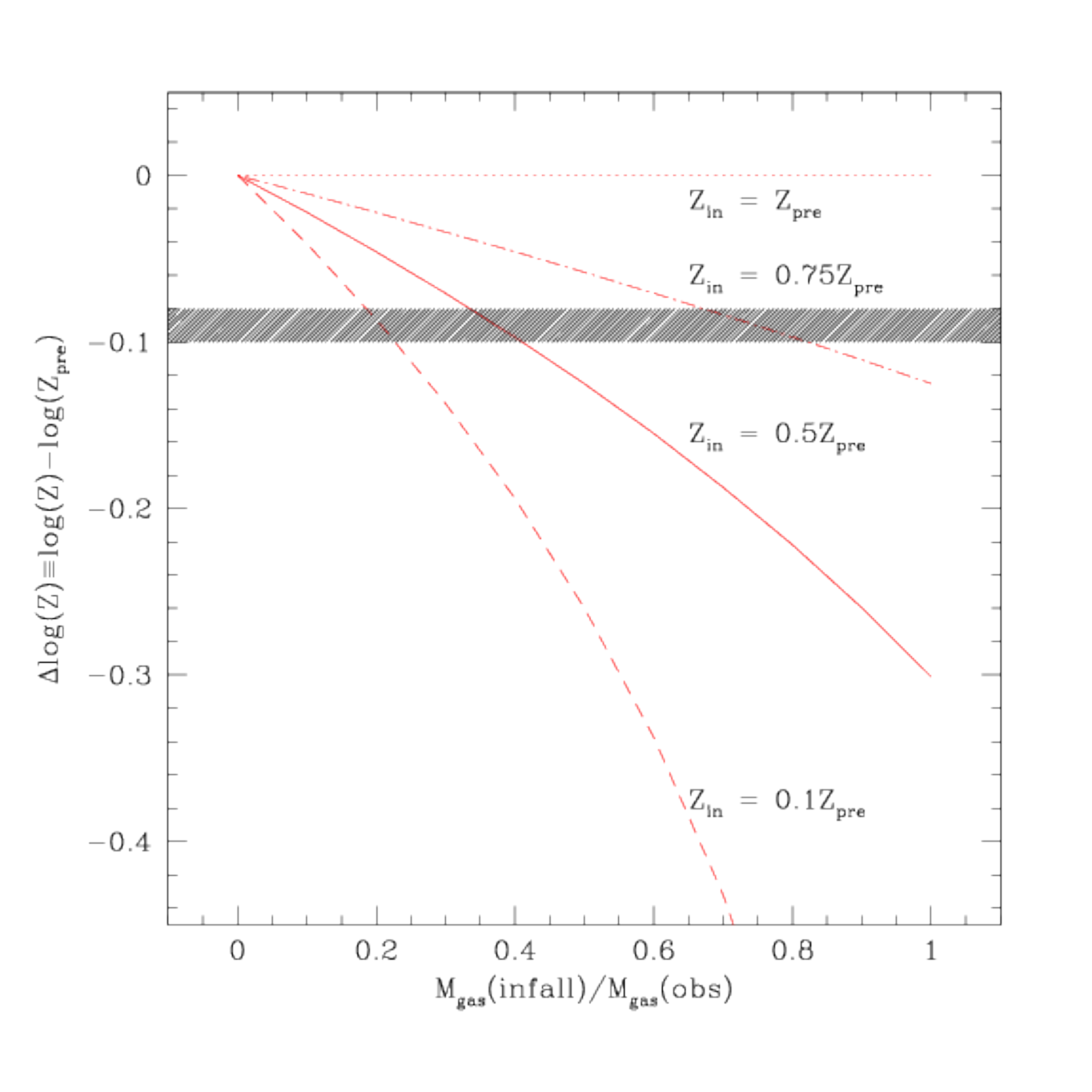}
\par\end{centering}
\caption{The effect of infall of metal-poor gas on the net metallicity. Curves are plotted for four values of abundance of infalling gas ($Z_{\rm in}$) in terms of the metallicity of the pre-collisional disk ($Z_{\rm pre}$). The observed deficit in the metallicity of the ring regions (the horizontal band) requires infall of the metal-poor gas, which constitutes $\sim$20--70\% of the observed gas mass, the exact value depending on the metallicity of the infalling gas.
}
\label{fig:metal_offset}
\end{figure}

In Fig.~\ref{fig:metal_offset}, we show the amount of metal-poor gas that is required to explain the observed downward offset in the abundances of the ring \hii\ regions with respect to that expected in the pre-collisional disk at the location of the outer ring. In this calculation, we assumed no metal-enrichment from the wave-induced SF. We infer that 50\% of the presently observed gas mass is of infall origin, if the metallicity of the infalling gas ($Z_{\rm in}$) is half of that of the pre-collisional disk ($Z_{\rm pre}$). The infall gas mass percentage could be as high as 80\% if $Z_{\rm in}=0.75\times Z_{\rm pre}$, or as low as 20\% if $Z_{\rm in}=0.1\times Z_{\rm pre}$. 

\subsection{Helium abundance}

The average He abundance for the 14 regions where we were able to estimate it is 12+log(He/H)$=10.90\pm0.03$. This value is close to the one found by \citet{Fosbury1977} of 12+log(He/H)$=10.86$ for the region we are labelling as \#99. The value we obtain is also close to the average value of \hii\ regions in M 33 reported by \citet{2016MNRAS.458.1866T} of 12+log(He/H)$=10.98\pm0.06$, the average value for \hii\ galaxies reported by \citet{2018MNRAS.478.5301F} of 12+log(He/H)$=10.94\pm0.04$,  as well as the median value of 12+log(He/H)$=10.9$ for galaxies with 12+log(O/H)$=8.0$ from the CALIFA survey \citep{Valerdi2021}, which implies that the enrichment of He has not been peculiar for the Cartwheel galaxy.

\section{Conclusions}

We have presented the nebular abundance analysis of 221 \hii\ regions identified in the external ring of the Cartwheel galaxy, as well as 40 regions inside the disk, using the spectroscopic data from the MUSE instrument.
We were able to determine the O, N, Fe and He abundances for a total of  20, 20, 17, and 9 regions, respectively, in the ring using the DM. 
For the regions in which  the temperature sensitive  \siiia\ line was not detected,
we have used the strong-line calibration method from \citet{2016MNRAS.457.3678P} to derive the O abundance in 152 and 24 regions in the external ring and the disk, respectively. These measurements show a systematic radial gradient corresponding to a slope of $-0.017\pm0.001\thinspace\rm{dex/kpc}$, which is consistent with those found in normal spirals \citep{2019A&A...623A.122P}. The observed O  abundances in the region between the nucleus and the outer ring follow the well-established local mass surface density-metallicity ($\Sigma_{*}$-$12+\log(\rm{O/H})$) relation  found for normal disk galaxies in the MANGA survey \citep{2016MNRAS.463.2513B}. 
This implies that the abundance of elements in the disk of the Cartwheel has little or no contribution from local enrichment
from the stars formed by the expanding wave in the past, or its dilution due to radial inflow of metal-poor gas. 
The empirically-derived O abundances drop by $\sim$0.1~dex at the ring, which suggests the presence of some amount of metal-poor gas in the ring. We do not find evidence for enrichment of He in the Cartwheel.
The abundance trends found in this work are in good agreement with the predictions of transfer of material (gas and recently formed stars) in the expanding wave obtained in recent simulation of the Cartwheel galaxy by \citet{Renaud2018} that takes into account gas-dynamical processes at parsec scales, or scenarios of the formation of ring galaxies in recent cosmological models by  \citet{2018MNRAS.481.2951E}. On the other hand, classical expanding density wave models fail to explain the observed abundance trends.

\section*{Acknowledgements}

This research has made use of the services of the ESO Science Archive Facility (program ID: 60.A-9333), PyNeb \citep{2015A&A...573A..42L}, Astropy,\footnote{\url{http://www.astropy.org}} a  community-developed core Python package for Astronomy \citep{2013A&A...558A..33A,2018AJ....156..123A}, APLpy,  an open-source plotting package for Python  \citep{2012ascl.soft08017R}, Astroquery, a package that contains a collection of tools to access online Astronomical data \citep{2019AJ....157...98G}, the MUSE Python Data Analysis  \citep[MPDAF]{2016ascl.soft11003B} and pyregion  ({\url{https://github.com/astropy/pyregion}}), a python module to parse ds9 region files. We thank the anonymous reviewer for the careful reading of our manuscript and the many insightful comments and suggestions. 
We also thank CONACyT for the research grant CB-A1-S-25070 (YDM).

\section{Data availability}

The fluxes of principal emission lines used in this work are available
in the article and in its online supplementary material. The reduced fits files
on which these data are based will be shared on reasonable request to the first
author. The ESO datacubes are in the public domain.

\appendix

\section{Supplementary material}

An extended version of Tab.~\ref{tab_abundances_direct} that contains measured fluxes and errors of all the emission lines used in this study is  available as a machine readable table for the 252 identified regions in the online version of the article. 

Here we list the description of each of the columns present in the MRT table:

{\scriptsize
\begin{enumerate}[leftmargin =5mm,labelwidth=4mm,label=\arabic*.]
\item Identification label.
\item Right Ascension (J2000).
\item Declination (J2000).
\item Electron temperature for the low ionization zone $T_{\rm{e}}^{\rm{low}}$.
\item Electron temperature for the low ionization zone error.
\item Electron temperature for the high ionization zone $T_{\rm{e}}^{\rm{high}}$.
\item Electron temperature for the high ionization zone error.
\item Electron temperature for the medium ionization zone $T_{\rm{e}}^{\rm{medium}}$.
\item Electron temperature for the medium ionization zone error.
\item Electron density $N_\mathrm{e}$  estimated from \sii$\lambda6717$/$\lambda6731$ lines.
\item Electron density error.
\item FeIII/H  abundance ratio.
\item FeIII/H  abundance ratio error.
\item OI/H abundance ratio.
\item OI/H abundance ratio error.
\item OII/H  abundance ratio.
\item OII/H  abundance ratio error.
\item OIII/H  abundance ratio.
\item OIII/H  abundance ratio error.
\item NII/H  abundance ratio.
\item NII/H  abundance ratio error.
\item HeI/H  abundance ratio.
\item HeI/H  abundance ratio error.
\item HeII/H  abundance ratio.
\item HeII/H  abundance ratio error.
\item Fe/H  abundance ratio.
\item Fe/H  abundance ratio error.
\item O/H  abundance ratio.
\item O/H  abundance ratio error.
\item N/H  abundance ratio.
\item N/H  abundance ratio error.
\item He/H  abundance ratio.
\item He/H  abundance ratio error.
\item Position Angle (PA) in the ring.
\item H$alpha$ Equivalent width.
\item $A_{\rm{V}}$ determined from the H$\alpha$ and H$\beta$ lines.
\item $A_{\rm{V}}$ error.
\item Galactrocentric radius $R_{\rm{gal}}$.
\item Oxygen abundance derived with the strong lines method from \citet{2016MNRAS.457.3678P}.
\item Oxygen abundance error derived with the strong lines method.
\item H$\beta$ Equivalent width.
\item SNR of H$\beta$ flux.
\item Attenuation-corrected flux of H$\beta$.
\item Attenuation-corrected flux error of H$\beta$.
\item $100\times\frac{F_{ \rm{[FeIII]4659A        }}}{F_{\rm{H\beta}}}                        $  ratio.
\item $100\times\frac{F_{ \rm{[FeIII]4659A        }}}{F_{\rm{H\beta}}}                   $       ratio error.
\item $100\times\frac{F_{ \rm{HeII4686A       }}}{F_{\rm{H\beta}}}                       $       ratio.
\item $100\times\frac{F_{ \rm{HeII4686A       }}}{F_{\rm{H\beta}}}                  $            ratio error.
\item $100\times\frac{F_{ \rm{[FeIII]4701A        }}}{F_{\rm{H\beta}}}                        $  ratio.
\item $100\times\frac{F_{ \rm{[FeIII]4701A        }}}{F_{\rm{H\beta}}}                   $       ratio error.
\item $100\times\frac{F_{ \rm{[ArIV]4711A        }}}{F_{\rm{H\beta}}}                        $   ratio.
\item $100\times\frac{F_{ \rm{[ArIV]4711A        }}}{F_{\rm{H\beta}}}                   $        ratio error.
\item $100\times\frac{F_{ \rm{[ArIV]4740A        }}}{F_{\rm{H\beta}}}                        $   ratio.
\item $100\times\frac{F_{ \rm{[ArIV]4740A        }}}{F_{\rm{H\beta}}}                   $        ratio error.
\item $100\times\frac{F_{ \rm{HeI4922A       }}}{F_{\rm{H\beta}}}                       $        ratio.
\item $100\times\frac{F_{ \rm{HeI4922A       }}}{F_{\rm{H\beta}}}                  $             ratio error.
\item $100\times\frac{F_{ \rm{[OIII]4959A         }}}{F_{\rm{H\beta}}}                         $ ratio.
\item $100\times\frac{F_{ \rm{[OIII]4959A         }}}{F_{\rm{H\beta}}}                    $      ratio error.
\item $100\times\frac{F_{ \rm{[FeIII]4986A        }}}{F_{\rm{H\beta}}}                        $  ratio.
\item $100\times\frac{F_{ \rm{[FeIII]4986A        }}}{F_{\rm{H\beta}}}                   $       ratio error.
\item $100\times\frac{F_{ \rm{[OIII]5007A         }}}{F_{\rm{H\beta}}}                         $ ratio.
\item $100\times\frac{F_{ \rm{[OIII]5007A         }}}{F_{\rm{H\beta}}}                    $      ratio error.
\item $100\times\frac{F_{ \rm{HeI5016A       }}}{F_{\rm{H\beta}}}                       $        ratio.
\item $100\times\frac{F_{ \rm{HeI5016A       }}}{F_{\rm{H\beta}}}                  $             ratio error.
\item $100\times\frac{F_{ \rm{HeI5048A       }}}{F_{\rm{H\beta}}}                       $        ratio.
\item $100\times\frac{F_{ \rm{HeI5048A       }}}{F_{\rm{H\beta}}}                  $             ratio error.
\item $100\times\frac{F_{ \rm{[FeIII]5270A        }}}{F_{\rm{H\beta}}}                        $  ratio.
\item $100\times\frac{F_{ \rm{[FeIII]5270A        }}}{F_{\rm{H\beta}}}                   $       ratio error.
\item $100\times\frac{F_{ \rm{[NII]5755A         }}}{F_{\rm{H\beta}}}                         $  ratio.
\item $100\times\frac{F_{ \rm{[NII]5755A         }}}{F_{\rm{H\beta}}}                    $       ratio error.
\item $100\times\frac{F_{ \rm{HeI5876A       }}}{F_{\rm{H\beta}}}                       $        ratio.
\item $100\times\frac{F_{ \rm{HeI5876A       }}}{F_{\rm{H\beta}}}                  $             ratio error.
\item $100\times\frac{F_{ \rm{[OI]6300A         }}}{F_{\rm{H\beta}}}                         $   ratio.
\item $100\times\frac{F_{ \rm{[OI]6300A         }}}{F_{\rm{H\beta}}}                    $        ratio error.
\item $100\times\frac{F_{ \rm{[SIII]6312A         }}}{F_{\rm{H\beta}}}                         $ ratio.
\item $100\times\frac{F_{ \rm{[SIII]6312A         }}}{F_{\rm{H\beta}}}                    $      ratio error.
\item $100\times\frac{F_{ \rm{[OI]6364A         }}}{F_{\rm{H\beta}}}                         $   ratio.
\item $100\times\frac{F_{ \rm{[OI]6364A         }}}{F_{\rm{H\beta}}}                    $        ratio error.
\item $100\times\frac{F_{ \rm{[NII]6548A         }}}{F_{\rm{H\beta}}}                         $  ratio.
\item $100\times\frac{F_{ \rm{[NII]6548A         }}}{F_{\rm{H\beta}}}                    $       ratio error.
\item $100\times\frac{F_{ \rm{H\alpha        }}}{F_{\rm{H\beta}}}                        $       ratio.
\item $100\times\frac{F_{ \rm{H\alpha        }}}{F_{\rm{H\beta}}}                   $            ratio error.
\item $100\times\frac{F_{ \rm{[NII]6584A         }}}{F_{\rm{H\beta}}}                         $  ratio.
\item $100\times\frac{F_{ \rm{[NII]6584A         }}}{F_{\rm{H\beta}}}                    $       ratio error.
\item $100\times\frac{F_{ \rm{HeI6678A       }}}{F_{\rm{H\beta}}}                       $        ratio.
\item $100\times\frac{F_{ \rm{HeI6678A       }}}{F_{\rm{H\beta}}}                  $             ratio error.
\item $100\times\frac{F_{ \rm{[SII]6716A         }}}{F_{\rm{H\beta}}}                         $  ratio.
\item $100\times\frac{F_{ \rm{[SII]6716A         }}}{F_{\rm{H\beta}}}                    $       ratio error.
\item $100\times\frac{F_{ \rm{[SII]6731A         }}}{F_{\rm{H\beta}}}                         $  ratio.
\item $100\times\frac{F_{ \rm{[SII]6731A         }}}{F_{\rm{H\beta}}}                    $       ratio error.
\item $100\times\frac{F_{ \rm{HeI7065A       }}}{F_{\rm{H\beta}}}                       $        ratio.
\item $100\times\frac{F_{ \rm{HeI7065A       }}}{F_{\rm{H\beta}}}                  $             ratio error.
\item $100\times\frac{F_{ \rm{HeI7281A       }}}{F_{\rm{H\beta}}}                       $        ratio.
\item $100\times\frac{F_{ \rm{HeI7281A       }}}{F_{\rm{H\beta}}}                  $             ratio error.
\item $100\times\frac{F_{ \rm{[OII]7319A         }}}{F_{\rm{H\beta}}}                         $  ratio.
\item $100\times\frac{F_{ \rm{[OII]7319A         }}}{F_{\rm{H\beta}}}                    $       ratio error.
\item $100\times\frac{F_{ \rm{[OII]7330A         }}}{F_{\rm{H\beta}}}                         $  ratio.
\item $100\times\frac{F_{ \rm{[OII]7330A         }}}{F_{\rm{H\beta}}}                    $       ratio error.
\item $100\times\frac{F_{ \rm{[ArIII]7751A        }}}{F_{\rm{H\beta}}}                        $  ratio.
\item $100\times\frac{F_{ \rm{[ArIII]7751A        }}}{F_{\rm{H\beta}}}                   $       ratio error.
\item $100\times\frac{F_{ \rm{[OI]8446A         }}}{F_{\rm{H\beta}}}                         $   ratio.
\item $100\times\frac{F_{ \rm{[OI]8446A         }}}{F_{\rm{H\beta}}}                    $        ratio error.
\item $100\times\frac{F_{ \rm{[SIII]9069A         }}}{F_{\rm{H\beta}}}                         $ ratio.
\item $100\times\frac{F_{ \rm{[SIII]9069A         }}}{F_{\rm{H\beta}}}                    $      ratio error.

\end{enumerate}
}


\begin{thebibliography}{99}

\bibitem[\protect\citeauthoryear{Appleton \& Marston}{1997}]{Appleton1997} 
Appleton P.~N., \& Marston, A.~P., 1997, AJ, 113, 201

\bibitem[\protect\citeauthoryear{Appleton \& Struck-Marcel}{1996}]{1996FCPh...16..111A}
Appleton, P.~N. \& Struck-Marcel, C. 2010, \fcp, 16, 111 

 \bibitem[\protect\citeauthoryear{Arellano-C{\'o}rdova \& Rodr{\'\i}guez}{2020}]{Karla2020}
Arellano-C{\'o}rdova K.~Z. \& Rodr{\'\i}guez M., 2020, MNRAS, 497, 672

\bibitem[\protect\citeauthoryear{Astropy Collaboration et al.}{2013}]{2013A&A...558A..33A} 
Astropy Collaboration et al., 2013, A\&A, 558, A33

\bibitem[\protect\citeauthoryear{Astropy Collaboration et al.}{2018}]{2018AJ....156..123A} 
Astropy Collaboration et al., 2018, AJ, 156, 123

\bibitem[\protect\citeauthoryear{Bacon et al.}{2010}]{Bacon2010}
Bacon, R. et al., 2010, \procspie, 7735, 773508


\bibitem[\protect\citeauthoryear{Bacon et al.}{2016}]{2016ascl.soft11003B} 
Bacon R., Piqueras L., Conseil S., et al., 2016, ascl.soft. ascl:1611.003



\bibitem[\protect\citeauthoryear{Bacon et al.}{2017}]{2017A&A...608A...1B}
Bacon R., Conseil S., Mary D., Brinchmann J. et al., 2017, A\&A, 608, A1

\bibitem[\protect\citeauthoryear{Baldwin, Phillips, \& Terlevich}{1981}]{1981PASP...93....5B}
Baldwin J.~A., Phillips M.~M. \& Terlevich R., 1981, PASP, 93, 5

\bibitem[\protect\citeauthoryear{Barrera-Ballesteros et al.}{2016}]{2016MNRAS.463.2513B}
Barrera-Ballesteros J.~K., Heckman T.~M., Zhu G.~B. et al., 2016, MNRAS, 463, 2513

\bibitem[\protect\citeauthoryear{Barrera-Ballesteros et al.}{2020}]{2020MNRAS.492.2651B}
Barrera-Ballesteros J.~K., Utomo D., Bolatto A.~D. et al., 2020, MNRAS, 492, 2651

\bibitem[\protect\citeauthoryear{Barway, Mayya \& Robleto-Or\'us}{2020}]{Barway2020}
Barway, S., Mayya, Y. D. \& Robleto-Or\'us, A. 2020, MNRAS, 497, 44


\bibitem[\protect\citeauthoryear{Belfiore et al.}{2017}]{2017MNRAS.469..151B}
Belfiore F., Maiolino R., Tremonti C. et al., 2017, MNRAS, 469, 151

\bibitem[\protect\citeauthoryear{Belfiore et al.}{2022}]{2022A&A...659A..26B}
Belfiore F., Santoro F., Groves B. et al., 2022, A\&A, 659, A26

\bibitem[\protect\citeauthoryear{Bell et al.}{2003}]{Bell2003}
Bell, E. F., McIntosh, D. H., Katz, N. et al., ApJSS, 2003, 149, 289

\bibitem[\protect\citeauthoryear{Bittner et al.}{2019}]{2019A&A...628A.117B} 
Bittner A., Falc{\'o}n-Barroso J., Nedelchev B. et al., 2019, A\&A, 628, A117

\bibitem[\protect\citeauthoryear{Bransford et al.}{1998}]{Bransford1998} 
Bransford M.~A., Appleton P.~N., Marston A.~P. et al., 1998, AJ, 116, 2757

\bibitem[\protect\citeauthoryear{Bresolin et al.}{2009}]{2009ApJ...700..309B} 
Bresolin F., Gieren W., Kudritzki R.-P. et al., 2009, ApJ, 700, 309

\bibitem[\protect\citeauthoryear{Brinchmann, Kunth \& Durret}{2008}]{Brinchmann2008} 
Brinchmann, J., Kunth, D. \& Durret, F., 2008, \aap, 485, 657-677

\bibitem[\protect\citeauthoryear{Cappellari}{2017}]{2017MNRAS.466..798C}
Cappellari M., 2017, MNRAS, 466, 798

\bibitem[\protect\citeauthoryear{Cardelli, Clayton \& Mathis}{1989}]{Cardelli1989}
Cardelli, J.~A., Clayton, G.~C. \& Mathis, J.~S. 1989, \apj, 345, 245

\bibitem[Crowther(2007)]{Crowther2007} 
Crowther, P.~A.\ 2007, \araa, 45, 177

\bibitem[\protect\citeauthoryear{Delgado-Inglada et al.}{2016}]{2016MNRAS.456.3855D} 
Delgado-Inglada G., Mesa-Delgado A., Garc{\'\i}a-Rojas J. et al., 2016, MNRAS, 456, 3855

\bibitem[\protect\citeauthoryear{Elagali et al.}{2018}]{2018MNRAS.481.2951E}
Elagali A., Lagos C.~D.~P., Wong O.~I. et al., 2018, MNRAS, 481, 2951

\bibitem[\protect\citeauthoryear{Eldridge et al.}{2017}]{Eldridge2017}
Eldridge, J.~J., Stanway, E.~R., Xiao, L. et al. 2017, PASA, 34, e058

\bibitem[\protect\citeauthoryear{Edmunds \& Pagel}{1978}]{1978MNRAS.185P..77E} 
Edmunds M.~G. \& Pagel B.~E.~J., 1978, MNRAS, 185, 77P

\bibitem[\protect\citeauthoryear{Espinosa-Ponce et al.}{2020}]{2020MNRAS.494.1622E}
Espinosa-Ponce C., S{\'a}nchez S.~F., Morisset C. et al., 2020, MNRAS, 494, 1622


\bibitem[\protect\citeauthoryear{Esteban et al.}{2009}]{2009ApJ...700..654E} 
Esteban C., Bresolin F., Peimbert M. et al., 2009, ApJ, 700, 654

\bibitem[\protect\citeauthoryear{Fern{\'a}ndez et al.}{2018}]{2018MNRAS.478.5301F} 
Fern{\'a}ndez V., Terlevich E., D{\'\i}az A.~I. et al., 2018, MNRAS, 478, 5301

\bibitem[\protect\citeauthoryear{Fosbury \& Hawarden}{1977}]{Fosbury1977}
Fosbury, R. A. E. \& Hawarden, T. G., 1977, MNRAS, 178, 473 

\bibitem[\protect\citeauthoryear{Garnett}{1990}]{1990ApJ...363..142G} 
Garnett D.~R., 1990, ApJ, 363, 142

\bibitem[\protect\citeauthoryear{Garnett}{1992}]{1992AJ....103.1330G} 
Garnett D.~R., 1992, AJ, 103, 1330

\bibitem[\protect\citeauthoryear{Ginsburg et al.}{2019}]{2019AJ....157...98G} 
Ginsburg A., Sip{\H{o}}cz B.~M., Brasseur C.~E. et al., 2019, AJ, 157, 98

\bibitem[\protect\citeauthoryear{Higdon}{1995}]{Higdon1995}
Higdon, J. L., 1995, ApJ, 455, 524 

\bibitem[\protect\citeauthoryear{Higdon}{1996}]{Higdon1996}
Higdon, J. L., 1996, ApJ, 467, 241

\bibitem[\protect\citeauthoryear{Higdon et al.}{2015}]{Higdon2015}
Higdon, J. L., Higdon, S. J. U., Mart\'{\i}n Ruiz et al., 2015, ApJ, 814, L1 


\bibitem[\protect\citeauthoryear{Hwang et al.}{2019}]{2019ApJ...872..144H}
Hwang H.-C., Barrera-Ballesteros J.~K., Heckman T.~M. et al., 2019, ApJ, 872, 144. 

\bibitem[\protect\citeauthoryear{Izotov et al.}{2006}]{2006A&A...448..955I} 
Izotov Y.~I., Stasi{\'n}ska G., Meynet G. et al., 2006, A\&A, 448, 955


 \bibitem[\protect\citeauthoryear{Izotov, Stasi{\'n}ska, \& Guseva}{2013}]{2013A&A...558A..57I}
Izotov Y.~I., Stasi{\'n}ska G. \& Guseva N.~G., 2013, A\&A, 558, A57



\bibitem[\protect\citeauthoryear{Kauffmann et al.}{2003}]{2003MNRAS.346.1055K}
Kauffmann G., Heckman T.~M., Tremonti C. et al., 2003, MNRAS, 346, 1055


\bibitem[\protect\citeauthoryear{Kehrig et al.}{2013}]{Kehrig2013} 
Kehrig, C., P{\'e}rez-Montero, E., V{\'\i}lchez, J.~M. et al. 2013, \mnras, 432, 2731-2745

\bibitem[\protect\citeauthoryear{Kennicutt, Bresolin, \& Garnett}{2003}]{2003ApJ...591..801K}
Kennicutt R.~C., Bresolin F. \& Garnett D.~R., 2003, ApJ, 591, 801


\bibitem[\protect\citeauthoryear{Kewley et al.}{2006}]{2006MNRAS.372..961K}
Kewley L.~J., Groves B., Kauffmann G. et al., 2006, MNRAS, 372, 961


\bibitem[\protect\citeauthoryear{Kewley, Nicholls \&  Sutherland}{2019}]{Kewley2019}
Kewley, L.~J., Nicholls, D.~C. \& Sutherland, R.~S., 2019, \araa, 57, 511-570


\bibitem[\protect\citeauthoryear{Kobulnicky et al.}{1997}]{Kobulnicky1997} 
Kobulnicky, H.~A., Skillman, E.~D., Roy, J.-R. et al., 1997, \apj, 477, 679-692

\bibitem[\protect\citeauthoryear{K\"oppen \& Hensler}{2005}]{Koppen2005}
K\"oppen, J \& Hensler, G. A\&A, 434, 531

\bibitem[\protect\citeauthoryear{Korchagin, Mayya, \& Vorobyov}{2001}]{Korchagin2001} 
Korchagin V., Mayya Y.~D. \& Vorobyov E., 2001, ApJ, 554, 281

\bibitem[\protect\citeauthoryear{Korchagin, Vorobyov \& Mayya}{1999}]{Korchagin1999} 
Korchagin V., Vorobyov E. \& Mayya Y.~D., 1999, ApJ, 522, 767

\bibitem[\protect\citeauthoryear{Kouroumpatzakis et al.}{2021}]{2021NGC922} 
Kouroumpatzakis, K., Zezas, A., Wolter, A. et al., 2021, MNRAS, 500, 962


\bibitem[\protect\citeauthoryear{Leitherer et al.}{1999}]{1999ApJS..123....3L}
Leitherer C., Schaerer D., Goldader J.~D. et al., 1999, ApJS, 123, 3

\bibitem[\protect\citeauthoryear{L{\'o}pez-S{\'a}nchez et al.}{2007}]{Lopez2007} 
L{\'o}pez-S{\'a}nchez, A.~R., Esteban, C., Garc{\'\i}a-Rojas, J. et al.,
2007, \apj, 656, 168-185

\bibitem[\protect\citeauthoryear{Luridiana, Morisset, \& Shaw}{2015}]{2015A&A...573A..42L} 
Luridiana V., Morisset C. \& Shaw R.~A., 2015, A\&A, 573, A42

\bibitem[\protect\citeauthoryear{Lynds \& Toomre}{1976}]{1976ApJ...209..382L}
Lynds, R. \& Toomre, A. 1976, \apj, 209, 382

\bibitem[\protect\citeauthoryear{Madore, Nelson \& Petrillo}{2009}]{Madore2009}
Madore, B. F., Nelson, E. \& Petrillo, K., 2009, \apjs, 181, 572-604

\bibitem[\protect\citeauthoryear{Maeder}{1992}]{Maeder1992} 
Maeder, A., 1992, \aap, 264, 105-120
%
\bibitem[\protect\citeauthoryear{Maiolino, R. \& Mannucci, F.}{2019}]{Maiolino2019} 
Maiolino, R. \& Mannucci, F., 2019, \aapr, 27, 3

\bibitem[\protect\citeauthoryear{Magrini, Gon{\c{c}}alves, \& Vajgel}{2017}]{2017MNRAS.464..739M}
Magrini L., Gon{\c{c}}alves D.~R. \& Vajgel B., 2017, MNRAS, 464, 739

\bibitem[\protect\citeauthoryear{Marcum, Appleton, \& Higdon}{1992}]{Marcum1992} 
Marcum P.~M., Appleton P.~N. \& Higdon J.~L., 1992, ApJ, 399, 57


\bibitem[\protect\citeauthoryear{Marino et al.}{2013}]{2013A&A...559A.114M}
Marino R.~A., Rosales-Ortega F.~F., S{\'a}nchez S.~F. et al., 2013, A\&A, 559, A114 



\bibitem[\protect\citeauthoryear{Marston \& Appleton}{1995}]{Marston1995} 
Marston, A.~P. \& Appleton P.~N., 1995, AJ, 109, 1002

\bibitem[\protect\citeauthoryear{Matteucci \& Greggio}{1986}]{Matteucci1986}
Matteucci, F. \& Greggio, L., 1986, A\&A, 154, 279

\bibitem[\protect\citeauthoryear{Mayya et al.}{2005}]{Mayya2005}
Mayya, Y.D., Bizyaev, D., Romano, R. et al., 2005, ApJ, 620, L35


\bibitem[\protect\citeauthoryear{McCall, Rybski, \& Shields}{1985}]{McCall1985} 
McCall M.~L., Rybski P.~M. \& Shields G.~A., 1985, ApJS, 57, 1



\bibitem[\protect\citeauthoryear{McKee \& Hollenbach}{1980}]{1980ARA&A..18..219M}
McKee C.~F. \& Hollenbach D.~J., 1980, ARA\&A, 18, 219. 

\bibitem[\protect\citeauthoryear{Menacho et al.}{2021}]{Menacho2021}
{Menacho}, V., {{\"O}stlin}, G. and {Bik}, A. et al., 2021, \mnras, 506, 1777

\bibitem[\protect\citeauthoryear{Osterbrock \& Ferland}{2006}]{Osterbrock2006}
Osterbrock, D.~E. \& Ferland, G.~J. 2006, in Astrophysics of Gaseous Nebulae and Active Galactic Nuclei (CA: University Science Books)

\bibitem[\protect\citeauthoryear{Ott}{2012}]{Ott2012}
Ott, T.\ 2012, QFitsView: FITS file viewer, ascl:1210.019

\bibitem[\protect\citeauthoryear{Makarov et al.}{2014}]{2014A&A...570A..13M}
Makarov D., Prugniel P., Terekhova N. et al., 2014, A\&A, 570, A13

\bibitem[\protect\citeauthoryear{Pagel, Terlevich \& Melnick}{1986}]{Pagel1986} 
Pagel, B.~E.~J., Terlevich, R.~J. \& Melnick, J., 1986, \pasp, 98, 1005-1008

\bibitem[\protect\citeauthoryear{Pagel et al.}{1992}]{Pagel1992} 
Pagel, B.~E.~J., Simonson, E.~A., Terlevich, R.~J. et al., 1992, \mnras, 255, 325-345


\bibitem[\protect\citeauthoryear{Pantelaki, Willson, \& Guzik}{1988}]{1988BAAS...20.1100P}
Pantelaki I., Willson L.~A. \& Guzik J., 1988, BAAS


\bibitem[\protect\citeauthoryear{Peimbert, Sarmiento, \& Fierro}{1991}]{1991PASP..103..815P}
Peimbert M., Sarmiento A. \& Fierro J., 1991, PASP, 103, 815



\bibitem[\protect\citeauthoryear{P\'erez-Montero}{2017}]{Perez-Montero2017}
P\'erez-Montero, E. 2017, PASP, 129, 043001



\bibitem[\protect\citeauthoryear{Pettini \& Pagel}{2004}]{2004MNRAS.348L..59P}
Pettini M. \& Pagel B.~E.~J., 2004, MNRAS, 348, L59



\bibitem[\protect\citeauthoryear{Pilyugin \& Grebel}{2016}]{2016MNRAS.457.3678P} 
Pilyugin L.~S. \& Grebel E.~K., 2016, MNRAS, 457, 3678

\bibitem[\protect\citeauthoryear{Pilyugin et al.}{2019}]{2019A&A...623A.122P} 
Pilyugin L.~S., Grebel E.~K., Zinchenko I.~A. et al., 2019, A\&A, 623, A122

\bibitem[\protect\citeauthoryear{Recchi et al.}{2004}]{Recchi2004} 
Recchi, S., Matteucci, F., D'Ercole, A. et al., 2004, A\&A, 426, 37

\bibitem[\protect\citeauthoryear{Recchi et al.}{2006}]{Recchi2006} 
Recchi, S., Hensler, G., Angeretti, L. et al., 2006, A\&A, 445, 875

\bibitem[\protect\citeauthoryear{Renaud et al.}{2018}]{Renaud2018} 
Renaud F., Athanassoula E., Amram P. et al., 2018, MNRAS, 473, 585

\bibitem[\protect\citeauthoryear{Renzini \& Voli}{1981}]{Renzini1981} 
Renzini, A. \& Voli, 1981, M. A\&A, 94, 175

\bibitem[\protect\citeauthoryear{Robitaille \& Bressert}{2012}]{2012ascl.soft08017R} 
Robitaille T., Bressert E., 2012, ascl.soft. ascl:1208.017

\bibitem[\protect\citeauthoryear{Rodr{\'\i}guez}{2002}]{2002A&A...389..556R} 
Rodr{\'\i}guez M., 2002, A\&A, 389, 556


\bibitem[\protect\citeauthoryear{Rosales-Ortega et al.}{2012}]{2012ApJ...756L..31R}
Rosales-Ortega F.~F., S{\'a}nchez S.~F., Iglesias-P{\'a}ramo J. et al., 2012, ApJL, 756, L31


\bibitem[\protect\citeauthoryear{Rupke, Kewley, \& Chien}{2010}]{2010ApJ...723.1255R}
Rupke D.~S.~N., Kewley L.~J. \& Chien L.-H., 2010, ApJ, 723, 1255


\bibitem[\protect\citeauthoryear{S{\'a}nchez et al.}{2014}]{2014A&A...563A..49S}
S{\'a}nchez S.~F., Rosales-Ortega F.~F., Iglesias-P{\'a}ramo J. et al., 2014, A\&A, 563, A49




\bibitem[\protect\citeauthoryear{S{\'a}nchez et al.}{2015}]{2015A&A...574A..47S}
S{\'a}nchez S.~F., P{\'e}rez E., Rosales-Ortega F.~F. et al., 2015, A\&A, 574, A47

\bibitem[\protect\citeauthoryear{S{\'a}nchez}{2020}]{Sanchez2020} 
S{\'a}nchez, S., 2020, \araa, 58, 99-155


\bibitem[\protect\citeauthoryear{S{\'a}nchez-Menguiano et al.}{2016}]{2016A&A...587A..70S}
S{\'a}nchez-Menguiano L., S{\'a}nchez S.~F., P{\'e}rez I. et al., 2016, A\&A, 587, A70




\bibitem[\protect\citeauthoryear{S{\'a}nchez-Menguiano et al.}{2018}]{2018A&A...609A.119S}
S{\'a}nchez-Menguiano L., S{\'a}nchez S.~F., P{\'e}rez I. et al., 2018, A\&A, 609, A119



\bibitem[\protect\citeauthoryear{S{\'a}nchez-Menguiano et al.}{2019}]{2019ApJ...882....9S}
S{\'a}nchez-Menguiano L., S{\'a}nchez Almeida J., Mu{\~n}oz-Tu{\~n}{\'o}n C. et al., 2019, ApJ, 882, 9

{\rm
\bibitem[\protect\citeauthoryear{Schlafly \& Finkbeiner}{2011}]{2011ApJ...737..103S} Schlafly E.~F., Finkbeiner D.~P., 2011, ApJ, 737, 103. 
}



\bibitem[\protect\citeauthoryear{Skrutskie et al.}{2006}]{2006AJ....131.1163S}
Skrutskie M.~F., Cutri R.~M., Stiening R. et al., 2006, AJ, 131, 1163


\bibitem[\protect\citeauthoryear{Struck}{2010}]{2010MNRAS.403.1516S}
Struck, C., 2010, \mnras, 403, 1516 

\bibitem[\protect\citeauthoryear{Tenorio-Tagle}{1996}]{Tenorio1996} 
Tenorio-Tagle, G., 1996, \aj, 111, 1641

\bibitem[\protect\citeauthoryear{Toribio San Cipriano et al.}{2016}]{2016MNRAS.458.1866T} 
Toribio San Cipriano L., Garc{\'\i}a-Rojas J., Esteban C. et al., 2016, MNRAS, 458, 1866

\bibitem[\protect\citeauthoryear{Tresse et al.}{1999}]{Tresse1999}
Tresse, L., Maddox, S., Loveday, J. \& Singleton, C. 1999, \mnras, 310, 262

\bibitem[\protect\citeauthoryear{Valerdi}{2021}]{Valerdi2021} 
Valerdi, M., et al., 2021, \mnras, 505, 5460-5467

\bibitem[\protect\citeauthoryear{Vazdekis et al.}{2010}]{2010MNRAS.404.1639V} 
Vazdekis A., S{\'a}nchez-Bl{\'a}zquez P., Falc{\'o}n-Barroso J. et al., 2010, MNRAS, 404, 1639

\bibitem[\protect\citeauthoryear{Vincenzo et al.}{2016}]{Vincenzo2016}
Vincenzo, F., Belfiore, F., Maiolino, R. et al., 2016, \mnras, 458, 3466

\bibitem[\protect\citeauthoryear{Villas-Costas \& Edmunds}{1993}]{Villas1993}
Villas-Costas, M. B. \& Edmunds, M. G., 1993, \mnras, 265, 199

\bibitem[\protect\citeauthoryear{Zurita \& Bresolin}{2012}]{2012MNRAS.427.1463Z} 
Zurita, A. \& Bresolin, F., 2012, MNRAS, 427, 1463


\bibitem[\protect\citeauthoryear{Zurita et al.}{2021}]{2021MNRAS.500.2359Z}
Zurita A., Florido E., Bresolin F. et al., 2021, MNRAS, 500, 2359

\end{thebibliography}
\end{document}